\documentclass{aa}
\usepackage{txfonts}
\usepackage{graphicx}
\usepackage{subfigure}
\usepackage{natbib}
\usepackage{amssymb}
\bibpunct{(}{)}{;}{a}{}{,}

\def\chandra{{\it Chandra~}}
\def\chandrak{{\it Chandra}}
\def\swift{{\it Swift~}}

\def\xmm{{XMM-Newton~}}
\def\xmmk{{XMM-Newton}}

\def\m31{{M~31}}
\def\msun{{$M_{\sun}$}}

\def\pe{PFF2005~}
\def\pz{PHS2007~}
\def\pek{PFF2005}
\def\pzk{PHS2007}

\newcommand{\nh}{\hbox{$N_{\rm H}$}~}
\newcommand{\hcm}[1]{$\times 10^{#1}$ cm$^{-2}$}

\newcommand{\ergs}[1]{$\times 10^{#1}$ \hbox{erg s$^{-1}$}}
\newcommand{\oergs}[1]{$10^{#1}$ erg s$^{-1}$}
\newcommand{\cts}[1]{$\times 10^{#1}$ ct s$^{-1}$}

\newcommand{\tpower}[1]{$\times 10^{#1}$}

\begin{document}

\title{X-ray monitoring of classical novae in the central region of \m31\\ I. June 2006 -- March 2007\thanks{Partly 
   based on observations with \xmmk, an ESA Science Mission with instruments and contributions directly funded by ESA Member States and NASA}}

\author{M.~Henze\inst{1}
	\and W.~Pietsch\inst{1}
	\and F.~Haberl\inst{1}
	\and M.~Hernanz\inst{2}
	\and G.~Sala\inst{3}
	\and M.~Della Valle\inst{4,5,6}
	\and D.~Hatzidimitriou\inst{7,8}
	\and A.~Rau\inst{1,9}
	\and D.H.~Hartmann\inst{10}
	\and J.~Greiner\inst{1}
	\and V.~Burwitz\inst{1}
	\and J.~Fliri\inst{11,12,13}
}

\institute{Max-Planck-Institut f\"ur extraterrestrische Physik, Giessenbachstra\ss e,
	D-85748 Garching, Germany\\
	email: mhenze@mpe.mpg.de
	\and Institut de Ci\`encies de l'Espai (CSIC-IEEC), Campus UAB, Fac. Ci\`encies, E-08193 Bellaterra, Spain	
	\and Departament de F\'isica i Enginyeria Nuclear, EUETIB (UPC-IEEC), Comte d'Urgell 187, 08036 Barcelona, Spain
	\and European Southern Observatory (ESO), D-85748 Garching, Germany
	\and INAF-Napoli, Osservatorio Astronomico di Capodimonte, Salita Moiariello 16, I-80131 Napoli, Italy
	\and International Centre for Relativistic Astrophysics, Piazzale della Repubblica 2, I-65122 Pescara, Italy
	\and Department of Astrophysics, Astronomy and Mechanics, Faculty of Physics, University of Athens, Panepistimiopolis, GR15784 Zografos, Athens, Greece
	\and Foundation for Research and Technology Hellas, IESL, Greece
	\and California Institute of Technology, Pasadena, CA 91125, USA
	\and Department of Physics and Astronomy, Clemson University, Clemson, SC 29634-0978, USA
	\and Universit\"atssternwarte M\"unchen, Scheinerstrasse, 81679 M\"unchen, Germany
	\and Instituto de Astrofisica de Canarias, E-38205 La Laguna, Tenerife, Spain
	\and Departamento de Astrofisica, Universidad de La Laguna, E-38205 La Laguna, Tenerife, Spain
}

\date{Received 2 April 2010 / Accepted 15 August 2010}

\abstract
{Classical novae (CNe) have recently been reported to represent the major class of supersoft X-ray sources (SSSs) in the central region of our neighbour galaxy \m31.}
{We carried out a dedicated monitoring of the \m31 central region with \xmm and \chandra in order to find SSS counterparts of CNe, determine the duration of their SSS phase and derive physical outburst parameters.}
{We systematically searched our data for X-ray counterparts of CNe and determined their X-ray light curves and spectral properties. Additionally, we determined luminosity upper limits for all novae from previous studies which are not detected anymore and for all CNe in our field of view with optical outbursts between May 2005 and March 2007.}
{We detected eight X-ray counterparts of CNe in \m31, four of which were not previously known. Seven sources can be classified as SSSs, one is a candidate SSS. Two SSSs are still visible more than nine years after the nova outburst, whereas two other nova counterparts show a short SSS phase of less than 150 days. Of the latter sources, M31N~2006-04a exhibits a short-time variable X-ray light curve with an apparent period of ($1.6\pm0.3$) h. This periodicity could indicate the binary period of the system. There is no X-ray detection for 23 out of 25 CNe which were within the field of view of our observations and had their outburst from about one year before the start of the monitoring until its end. From the 14 SSS nova counterparts known from previous studies, ten are not detected anymore. Additionally, we found four SSSs in our \xmm data without a nova counterpart, one of which is a new source.}
{Out of eleven SSSs detected in our monitoring, seven are counterparts of CNe. We therefore confirm the earlier finding that CNe are the major class of SSSs in the central region of \m31. We use the measured SSS turn-on and turn-off times to estimate the mass ejected in the nova outburst and the mass burned on the white dwarf. Classical novae with short SSS phases seem to be an important contributor to the overall population.}

\keywords{Galaxies: individual: \m31 -- novae, cataclysmic variables -- X-rays: binaries -- stars: individual: M31N~1996-08b, M31N~1997-11a, M31N~2001-01a, M31N~2001-10a, M31N~2004-05b, M31N~2005-02a, M31N~2006-04a, M31N~2006-06a}

\titlerunning{X-ray monitoring of classical novae in \m31 from June 2006 to March 2007}

\maketitle

%
%
\section{Introduction}
%
This is the first of two papers analysing recent X-ray monitoring campaigns for classical novae in the central region of our neighbour galaxy \m31. Here we present the results of the first campaign from June 2006 to March 2007.

\begin{figure*}[t]
	\resizebox{\hsize}{!}{\includegraphics[angle=0]{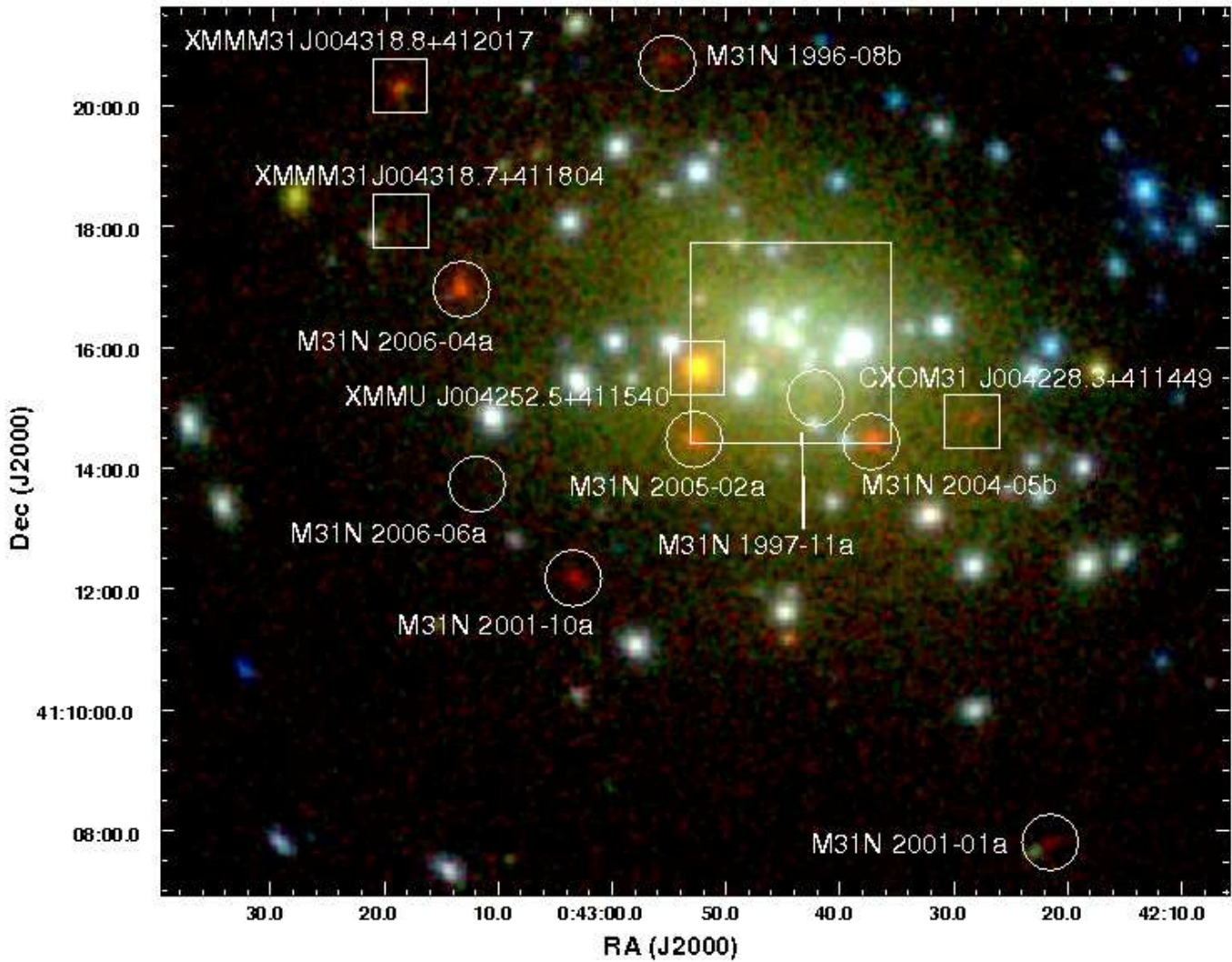}}
	\caption{Logarithmically-scaled, three colour \xmm EPIC image of the central area of \m31 combining PN, MOS1 and MOS2 data of all five observations of the campaign. Red, green and blue show the (0.2 -- 0.5) keV, (0.5 -- 1.0) keV and (1.0 -- 2.0) keV bands. The SSSs show up deeply red. The data in each colour band were binned in 2\arcsec x 2\arcsec pixels and smoothed using a Gaussian of FWHM 5\arcsec. The counterparts of optical novae detected in this work are marked with white circles. For M31N~1997-11a and M31N~2006-06a only the positions are designated, since they are not visible is this image but are detected in \chandra images. The four non-nova SSSs detected in this work are marked with white boxes. The large white box includes the central region of \m31, which is shown as a \chandra composite in Fig.\,\ref{fig:chandra}.}
	\label{fig:xmm}
\end{figure*}

Classical novae (CNe) originate from thermonuclear explosions on the surface of white 
dwarfs (WDs) in cataclysmic binaries. These explosions result from the transfer of matter 
from the companion star to the WD. The transferred hydrogen-rich matter accumulates 
on the surface of the WD until hydrogen ignition starts a thermonuclear runaway 
in the degenerate matter of the WD envelope. The resulting expansion of the hot 
envelope causes the brightness of the WD to rise by a typical outburst amplitude of $\sim$ 12 magnitudes 
within a few days, and mass to be ejected at high velocities \citep[see][and references 
therein]{2005ASPC..330..265H,1995cvs..book.....W}. However, a fraction of the hot 
envelope can remain in steady hydrogen burning on the surface of the WD 
\citep{1974ApJS...28..247S,2005A&A...439.1061S}, powering a supersoft X-ray source 
(SSS) that can be observed directly once the ejected envelope becomes sufficiently 
transparent \citep{1989clno.conf...39S,2002AIPC..637..345K}. In this paper, we define the term ``turn-on of the SSS'' from an observational point of view as the time when the SSS becomes visible to us, owing to the decreasing opacity of the ejected material. This is not the same as the actual turn-on of the H-burning on the WD surface at the beginning of the thermonuclear runaway. The turn-off of the SSS, on the other hand, does not suffer from extinction effects and clearly marks the actual turn-off of the H-burning in the WD atmosphere and the disappearance of the SSS.

The class of SSSs was first characterised on the basis of ROSAT observations \citep{1991Natur.349..579T,1991A&A...246L..17G}. These sources show extremely soft X-ray spectra, with little or no emission at energies above 1 keV, which can be described by equivalent blackbody temperatures of $\sim$15-80 eV \citep[see][and references therein]{1997ARA&A..35...69K}. It is well known that blackbody fits to SSS spectra are \textit{not} a physically correct model. These fits produce in general too high values of \nh and too low temperatures, resulting in overestimated luminosities \citep[see e.g.][and references therein]{1991A&A...246L..17G, 1997ARA&A..35...69K}. A more realistic approach would be to use stellar atmosphere models that assume non-local thermodynamic equilibrium (NLTE) \citep[see e.g.][and references therein]{2010AN....331..175V}. These models are still in development, but have been tested on bright Galactic CNe with promising results for static atmospheres \citep[see e.g.][]{2008ApJ...673.1067N} and expanding atmospheres \citep[see e.g.][]{2005A&A...431..321P}. However, these models are more justified when interpreting high spectral resolution observations. Supersoft X-ray sources in \m31 \citep[distance 780 kpc;][]{1998AJ....115.1916H,1998ApJ...503L.131S} are not bright enough to be observed with X-ray grating spectrometers. Furthermore, in the \m31 central region the source density is too high and the diffuse emission too strong to successfully apply grating spectroscopy for individual sources. Owing to the low resolution and low signal-to-noise ratio of the \xmm EPIC PN spectra analysed in this work, we decided to apply blackbody fits. In this way, we use the blackbody temperature as a simple parametrisation of the spectrum and can compare our results with earlier work \citep[e.g.][]{2005A&A...442..879P,2007A&A...465..375P}, while keeping in mind the biases outlined above.

The duration of the SSS phase in CNe is related to the amount of H-rich matter that is 
{\it not} ejected and on the luminosity of the WD. More massive WDs need to accrete less matter to initiate the 
thermonuclear runaway, because of their higher surface gravity \citep{1998ApJ...494..680J}. Together with the fact that their stronger gravitational potential prevents more massive ejecta, this is causing more massive WDs to eject less matter in the nova outburst \citep{2005ApJ...623..398Y}. This theoretical result is confirmed by observations which found less massive ($\sim 2$\tpower{-5} $M_{\sun}$) ejecta associated with the brightest (very fast declining) novae. For slow novae the mass of the ejecta is about 10 times larger \citep{2002A&A...390..155D}.
In general, more massive WDs retain less accreted mass after the explosion, although this also 
depends on the accretion rate, and reach higher luminosities \citep{2005ApJ...623..398Y}. Thus, the 
duration of the SSS state is inversely related to the mass of the WD, for a given hydrogen-mass fraction in the remaining envelope. However, the larger the hydrogen content, the longer the duration of the SSS state for a given WD mass \citep[see][]{2005A&A...439.1061S,1998ApJ...503..381T,2006ApJS..167...59H}.

In turn, the transparency requirement mentioned above implies that the turn-on time is determined by the fraction of mass 
ejected in the outburst. X-ray light curves therefore provide important clues to the physics of the nova outburst, 
addressing the key question of whether a WD accumulates matter over time to become a potential
progenitor for a type Ia supernova (SN-Ia). The duration of the SSS state provides the only 
direct indicator of the post-outburst envelope mass in CNe. For massive WDs, the expected SSS 
phase is very short ($<$ 100 d) \citep{2005A&A...439.1061S,1998ApJ...503..381T}.

Recurrent novae (RNe) are classified by the observational fact that they have had more than one recorded nova outburst. These objects are very good candidates for SN-Ia progenitors in the single degenerate scenario, in which a carbon-oxygen (CO) WD accretes matter from a non-degenerate companion \citep{1973ApJ...186.1007W,1982ApJ...253..798N}, as RNe are believed to contain massive WDs. Alternatively, two WDs that merge can cause a SN-Ia if one of them is a CO WD \citep[= double degenerate scenario][]{1984ApJ...277..355W,1984ApJS...54..335I}. However, because nova outbursts occur on accreting WDs, only the single degenerate scenario is important in the context of this work. Optical surveys show that RNe are rare relative to the observed SN-Ia rate \citep{1996ApJ...473..240D}. But hypothetical previous outbursts may have been missed, and then these objects might have been classified as CNe in optical surveys. Interesting cases are CNe with very short SSS phases because they indicate a high WD mass. These objects are potentially capable of frequent outbursts as in RNe.

Owing to its proximity to the Galaxy and its moderate Galactic foreground absorption \citep[\nh $\sim 6.7$ \hcm{20},][]{1992ApJS...79...77S}, \m31 is a unique target for CN surveys, which have been carried out starting with the seminal work of \citet{1929ApJ....69..103H} \citep[see also][and references therein]{2008A&A...477...67H,2001ApJ...563..749S}. However, only recently nova monitoring programmes for \m31 were established that include fast data analysis and therefore provide the possibility to conduct follow-up spectroscopy \citep[see e.g.][]{2007A&A...464.1075H} and to confirm and classify CNe within the system of \citet{1992AJ....104..725W}.

The advantages and disadvantages of X-ray surveys for CNe in \m31 compared to Galactic surveys \citep[e.g.][]{2007ApJ...663..505N,2001A&A...373..542O} are similar to those of optical surveys. In our galaxy, the investigation of the whole nova population is hampered by the large area (namely the whole sky) to be scrutinised and by our unfavourable position close to the Galactic Plane. As \citet{2007ApJ...663..505N} pointed out, the detectability of a SSS is highly dependent on its foreground absorption, which is affecting the supersoft X-rays much stronger than harder photons. Note that \citet{2001A&A...373..542O} analysed 350 archival ROSAT observations and found SSS emission for only three Galactic novae. Galactic novae and SSSs can of course be studied in much greater detail than is possible for \m31 objects, simply because they are closer to us. But they have to be observed one by one. Furthermore, determining the actual distance to a Galactic CN is not trivial. Observations of \m31 on the other hand yield light curves of many CNe simultaneously and all of these objects are effectively at the same, well-known distance. While Galactic sources therefore are the ideal targets to examine the SSS emission of individual novae in detail, observations of \m31 allow us to study the ``big picture'' and provide insight into the CN population of a large spiral galaxy.

\citet[][hereafter \pek]{2005A&A...442..879P} correlated CNe catalogues of \m31 with X-ray catalogues from ROSAT, \xmmk, and \chandra and found that CNe represent the major class of SSSs in this galaxy. They state that more than 60\% of the SSSs in the \m31 central area can be identified with novae. After that, \citet[][hereafter \pzk]{2007A&A...465..375P} presented a search for X-ray counterparts of CNe in \m31 based on archival \chandra and \xmm observations and concluded that the number of CNe detected as SSSs is much higher than previously estimated ($>30$\%). This work is a follow-up of \pe and \pz and presents the results of the first dedicated monitoring of CNe in the \m31 central region with \xmm and \chandrak. The target was densely monitored with nine observations from June 2006 to March 2007. Additionally, we make use of nine archival \chandra ACIS-I and two archival \swift XRT observations of the \m31 centre during that time. Altogether, our data set consists of all \xmmk, \chandra and \swift observations performed after the last observation of \pz until March 2007. A second paper (Henze et al., in prep.) will study the monitoring campaigns in 2007/8 and 2008/9 and discuss global properties of CNe as SSSs in \m31. In Sect.\,\ref{sec:obs}, we describe our X-ray observations and data analysis. Results are presented in Sect.\,\ref{sec:results} and discussed in Sect.\,\ref{sec:discuss}.

%
%
\section{Observations and data analysis}
\label{sec:obs}
%

\begin{figure}
	\vspace{0.6cm}
	\resizebox{\hsize}{!}{\includegraphics[angle=0]{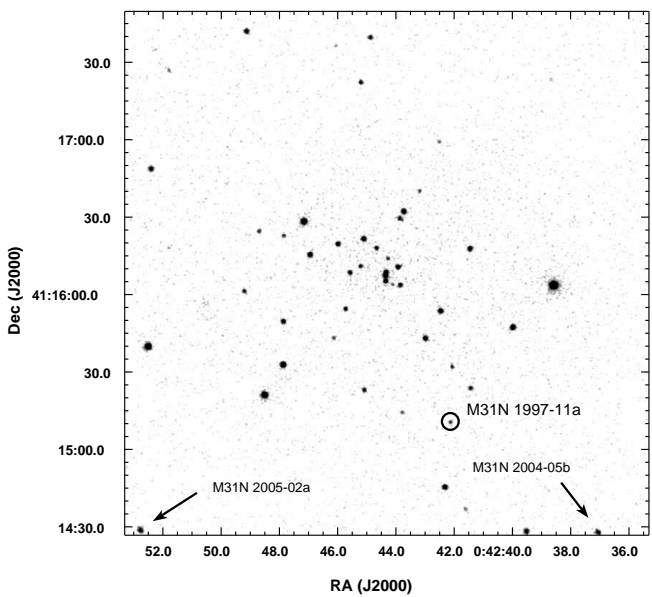}}
	\caption{Logarithmically-scaled \chandra HRC-I image of the innermost $3\farcm3 \times 3\farcm3$ of \m31 combining all four observations of the campaign. The image has not been binned (HRC electronic pixel size = $0\farcs13$) but has been smoothed with a Gaussian of FWHM $0\,\farcs5$. The X-ray counterpart of nova M31N 1997-11a is marked with a black circle. On the lower edges of the field of view the X-ray counterparts of novae M31N 2005-02a and M31N 2004-05b can be seen.}
	\label{fig:chandra}
\end{figure}

The X-ray data used in this work were obtained in a joint \xmmk/\chandra monitoring programme (PI: W. Pietsch)\footnote{http://www.mpe.mpg.de/$\sim$m31novae/xray/index.php}. We monitored the \m31 central region with five \xmm European Photon Imaging Camera (EPIC) and four \chandra High-Resolution Camera Imaging Detector (HRC-I) observations that were distributed between June 2006 and March 2007. The dates, observation identifications (ObsIDs) and dead-time corrected exposure times of the individual observations are given in Table\,\ref{table:obs}. The field of view of the monitoring is shown in an \xmm EPIC colour image in Fig.\,\ref{fig:xmm}.

%
\begin{table*}
\begin{center}
\caption[]{Observations of the \m31 monitoring.}
\begin{tabular}{lrrrrrrrrr}
\hline\noalign{\smallskip}
\hline\noalign{\smallskip}
\multicolumn{1}{l}{Telescope/Instrument} & \multicolumn{1}{r}{ObsID} & \multicolumn{4}{c}{Exposure Time$^a$ [ks]}
& \multicolumn{1}{c}{Start Date$^b$} & \multicolumn{1}{c}{JD$^b$} & \multicolumn{2}{c}{Offset$^c$}\\
\noalign{\smallskip}
 & & \multicolumn{1}{r}{PN} & \multicolumn{1}{r}{MOS1} & \multicolumn{1}{r}{MOS2} 
& \multicolumn{1}{r}{HRC-I} & \multicolumn{1}{c}{[UT]} & \multicolumn{1}{l}{2\,450\,000+} 
& \multicolumn{1}{r}{RA [$\arcsec$]} & \multicolumn{1}{r}{Dec [$\arcsec$]}\\
\noalign{\smallskip}\hline\noalign{\smallskip}
	\chandra HRC-I & 7283 & & & & 19.9 & 2006-06-05.29 & 3891.79 & 0.2 & 0.4\\
	\xmm EPIC & 0405320501 & 10.8 & 13.0 & 13.5 & & 2006-07-02.61 & 3919.11 & 2.1 & 1.8\\
	\xmm EPIC & 0405320601 & 9.8 & 16.3 & 12.0 & & 2006-08-09.51 & 3957.02 & 1.6 & 1.6\\
	\chandra HRC-I & 7284 & & & & 20.0 & 2006-09-30.89 & 4009.39 & -0.3 & 0.4\\
	\chandra HRC-I & 7285 & & & & 18.5 & 2006-11-13.30 & 4052.80 & -0.4 & 0.0\\
	\xmm EPIC & 0405320701 & 12.3 & 15.4 & 15.4 & & 2006-12-31.60 & 4101.10 & 0.8 & 0.9\\ 
	\xmm EPIC & 0405320801 & 9.9 & 13.2 & 13.4 & & 2007-01-16.49 & 4116.99 & 0.0 & 0.5\\ 
	\xmm EPIC & 0405320901 & 12.5 & 16.3 & 16.3 & & 2007-02-05.15 & 4136.66 & -0.4 & 2.0\\
	\chandra HRC-I & 7286 & & & & 18.7 & 2007-03-11.61 & 4171.11 & -0.3 & -0.1\\
\noalign{\smallskip} \hline
\end{tabular}
\label{table:obs}
\end{center}
Notes:\hspace{0.3cm} $^a $: Dead-time corrected; for \xmm EPIC after screening for high background and PN, MOS1, MOS2.\\
\hspace*{1.1cm} $^b $: Start date of observations; for \xmm EPIC the PN start date was used.\\
\hspace*{1.1cm} $^c $: Offset of image WCS to the WCS of the catalogue by \citet{2002ApJ...578..114K}.\\
\end{table*}

In the \xmm observations, the EPIC PN and MOS instruments were operated in the full frame mode. We used the thin filter for PN and the medium filter for MOS. The \xmm data processing and analysis was carried out in the context of XMMSAS v8.0 \citep[\xmm Science Analysis System;][]{2004ASPC..314..759G}\footnote{http://xmm.esac.esa.int/external/xmm\_data\_analysis/}. However, the XMMSAS packages used in this work did not change significantly in newer versions. While source detection was performed on the EPIC PN and MOS images simultaneously, we only used PN data for extracting source spectra because of the better sensitivity of PN in the soft energy band.

For \xmm source detection, we rejected times with high background. The EPIC PN data were first processed with the XMMSAS task \texttt{epreject}, which corrects shifts in the energy scale of specific pixels that were caused by high-energy particles hitting the detector during the calculation of the offset map at the beginning of each observation. Additionally, the PN data were carefully screened in order to exclude bad pixels and bad columns in individual CCDs. From the screened PN, MOS1 and MOS2 event files we created images for the energy bands 0.2 -- 0.5, 0.5 -- 1.0, 1.0 -- 2.0, 2.0 -- 4.5, and 4.5 -- 12.0 keV. These images were binned to a pixel size of $2\arcsec \times 2\arcsec$. For EPIC PN, we selected single and double-pixel events (PATTERN $\leq$ 4) for image creation for all bands except for the 0.2 -- 0.5 keV band, where we used only singles-pixel events (PATTERN = 0). For EPIC MOS, we used single to quadruple-pixel events (PATTERN $\leq$ 12). Background maps were created for every image using the XMMSAS task \texttt{esplinemap}. The 15 images (five energy bands per detector) were processed simultaneously, using the XMMSAS commands \texttt{eboxdetect} and \texttt{emldetect}, to detect sources in the images and perform astrometry and photometry.

We reduced the \chandra HRC-I observations with the CIAO v3.4 software package \citep[Chandra Interactive Analysis of Observations;][]{2006SPIE.6270E..60F}\footnote{http://cxc.harvard.edu/ciao/}, starting from the level 2 event files. In order to account for the \chandra PSF significantly degrading with increasing off-axis angle we produced five images covering the central 200\arcsec, 400\arcsec, 800\arcsec, 1400\arcsec regions and the full field of view with a resolution of 1, 2, 4, 10, and 16 HRC electronic pixels ($0\farcs13$), respectively. The source detection was done with the CIAO tool \texttt{wavedetect} on the individual images. An adapted version of the XMMSAS tool \texttt{emldetect} was used to estimate background-corrected and exposure-corrected fluxes and count rates for the detected sources. Finally, we merged the individual detection lists to create a source list for the whole observation.

For estimating count rate upper limits for non-detected X-ray counterparts of novae, we added an artificial detection at the position of the source to the \texttt{eboxdetect} list of the observation. This new list was used as input for \texttt{emldetect} (with fixed positions and likelihood threshold of zero) to derive the observed count rate, or an upper limit, for all objects in the list. We applied this procedure to \xmm and \chandra data.

We computed intrinsic luminosities and 3$\sigma$ upper limits in the 0.2 -- 1.0 keV band starting from the 0.2 -- 0.5 and 0.5 -- 1.0 keV count rates or upper limits in \xmm EPIC and from the full 0.2 -- 10.0 keV count rates in \chandra HRC-I. Count rates and fluxes in the two individual bands were computed separately and added up to the 0.2 -- 1.0 keV band values. In order to compare our results to the earlier work of \pzk, luminosities given in all tables were computed assuming a typical 50 eV blackbody spectrum with Galactic foreground absorption (\nh $\sim 6.7$ \hcm{20}). Table\,\ref{tab:ecf} gives energy conversion factors (ECF = (count rate)/flux) for a range of blackbody temperatures and the different instruments, clearly showing that the ECFs depend strongly on the temperature of the spectrum. 

Additionally, we give for every SSS spectrum the individual best-fit parameters and the resulting formal luminosities in the text. Note that some of these luminosities exceed the Eddington limit for even a very massive WD with $M_{WD} \gtrsim 1.3M_{\sun}$ ($L_{\hbox{\rm Edd}} \sim 1.7 \times 10^{38}$). This is because of the general effect of overestimated luminosities in blackbody models and because that additional absorption in \m31 would drastically change the observed count rate (see also \pzk), in particular for very soft X-ray sources. All luminosities assume a distance to \m31 of 780 kpc.

%
\begin{table}
\begin{center}
\caption[]{Energy conversion factors (ECF = (count rate)/flux) for blackbody models with different temperatures, including a Galactic foreground absorption of 6.7\hcm{20}, for the instruments used in this work.}
\begin{tabular}{lrrrrr}
\hline\noalign{\smallskip}
\hline\noalign{\smallskip}
\multicolumn{1}{l}{Detector} & \multicolumn{1}{c}{Band} &\multicolumn{1}{c}{20 eV} &\multicolumn{1}{c}{30 eV} 
 & \multicolumn{1}{c}{40 eV} &\multicolumn{1}{c}{50 eV}  \\ 
\noalign{\smallskip}
 & \multicolumn{1}{c}{[eV]} & \multicolumn{4}{c}{($10^{10}$ ct cm$^{2}$ erg$^{-1}$)} \\
\noalign{\smallskip}\hline\noalign{\smallskip}
EPIC PN & 0.2--0.5 & 1.12 & 3.28 & 6.69 & 10.8\\
 & 0.5--1.0 & 113 & 58.9 & 53.1 & 52.1\\
EPIC MOS & 0.2--0.5 & 0.07 & 0.33 & 0.86 & 1.60\\
 & 0.5--1.0 & 15.7 & 6.62 & 6.89 & 7.40\\
HRC-I & 0.2--10.0 & 0.40 & 0.66 & 1.19 & 1.94\\
\noalign{\smallskip}
\hline
\noalign{\smallskip}
\end{tabular}
\label{tab:ecf}
\end{center}
\end{table}

To improve the astrometric accuracy of our measurements we applied a pointing offset correction according to the X-ray source catalogue of \citet{2002ApJ...578..114K}. This catalogue was calibrated astrometrically using the Two Micron All Sky Survey \citep[2MASS,][]{2003tmc..book.....C}, and according to the author the average displacement is $0\,\farcs15$ to 2MASS source. The corrected offset values are given in Table\,\ref{table:obs}.

The astrometrically calibrated objects from each observation were correlated with the updated catalogue of optical novae in \m31 by \citet{2007A&A...465..375P}\footnote{http://www.mpe.mpg.de/$\sim$m31novae/opt/m31/index.php}. X-ray sources were identified as candidate counterparts of CNe if their positions matched within a certain search radius. This search radius was determined by the positional error of the CN as given in the catalogue and the 3$\sigma$ error of the X-ray sources in the individual observation. We constrained the search for X-ray counterparts to CNe with outbursts after 1995. This was done to suppress spurious identifications, since the positions of earlier CNe are often not as well determined. In addition, one would only expect the start of the SSS phase later than 10 years after outburst for CNe under extreme conditions \citep[low WD mass and core material with close to solar abundance, see][]{2006ApJS..167...59H}. The candidate counterparts automatically found were visually inspected. We rejected candidates that turned out to be mis-identifications of known persistent hard sources in the field.

In order to increase the signal-to-noise ratio of low-luminosity sources, we merged all the \xmm data of the monitoring campaign for each detector (EPIC PN, MOS1, MOS2). We first re-adjusted the WCS (world coordinate system) of the screened event files by computing the offset between the original WCS and the catalogue we used for astrometrical calibration. The pointing offsets are given in Table\,\ref{table:obs} and were corrected in the attitude file of the observation. Then we shifted the pixel coordinates of the event files to a common reference point using the XMMSAS command \texttt{attcalc}. Images for the three detectors and five energy bands given above were created for every observation (see Table\,\ref{table:obs}) from the WCS corrected event files. Then, for each energy band and detector, we added up the five images from the individual observations to one merged image. Background map creation and object detection on the merged images was done analogously to the individual observations. Furthermore, the EPIC PN event files were stacked using the XMMSAS task \texttt{merge}, in order to extract spectra of low-luminosity CN counterparts. An \xmm EPIC colour image created from the merged images is presented in Fig.\,\ref{fig:xmm}.

For \chandra HRC-I data, we applied a similar strategy. We again computed the position offsets to the catalogue (see Table\,\ref{table:obs}) and applied them to correct the aspect solution file with the CIAO task \texttt{wcs\_update}. The updated aspect solution was then used to re-project the level 2 event file coordinates to the catalogue system (CIAO procedure \texttt{reproject\_events}). We created images and exposure maps for each observation and merged them using FTOOLS tasks. In Fig.\,\ref{fig:chandra} we show the merged image of the very centre of \m31. The merged data were processed as described above for the individual observations. Most luminosity upper limits for undetected CNe were estimated with the merged \xmm and \chandra data.

Additionally, during the time of our monitoring there were nine observations of the \m31 central region with \chandra ACIS-I and two observations with the \swift X-ray Telescope (XRT). Details are given in Table\,\ref{tab:acis}. However, only two of the CN counterparts discussed in this work are visible in these archival data. The source M31N~2004-05b (see Sect.\,\ref{sec:res_known}) can be seen as a faint object in both \swift observations, whereas M31N~2005-02a (see Sect.\,\ref{sec:res_new}) is detected with about seven counts in the 23 ks observation 7064. No other known nova is detected.

%
\begin{table}
\begin{center}
\caption[]{Archival \chandra ACIS-I and \swift XRT observations of the \m31 central region.}
\begin{tabular}{lrrr}
\hline\noalign{\smallskip}
\hline\noalign{\smallskip}
Telescope/Instrument & \multicolumn{1}{r}{ObsID} & \multicolumn{1}{r}{$t_{\mbox{exp}}^{\,\,a}$}
& \multicolumn{1}{c}{Start Date}\\
\noalign{\smallskip}
 & & \multicolumn{1}{c}{[ks]} & \multicolumn{1}{c}{[UT]} \\
\noalign{\smallskip}\hline\noalign{\smallskip}
\chandra ACIS-I & 7136 & 4.0 & 2006-01-06.84\\
\chandra ACIS-I & 7137 & 4.0 & 2006-05-26.18\\
\chandra ACIS-I & 7138 & 4.1 & 2006-06-09.68\\
\chandra ACIS-I & 7139 & 4.0 & 2006-07-31.02\\
\swift XRT & 00030802001 & 5.9 & 2006-09-01.75\\
\swift XRT & 00030804001 & 5.9 & 2006-09-11.73\\
\chandra ACIS-I & 7140 & 4.1 & 2006-09-24.76\\
\chandra ACIS-I & 7064 & 23.2 & 2006-12-04.88\\
\chandra ACIS-I & 8183 & 4.0 & 2007-01-14.86\\
\chandra ACIS-I & 8184 & 4.1 & 2007-02-14.08\\
\chandra ACIS-I & 8185 & 4.0 & 2007-03-10.25\\
\noalign{\smallskip} \hline
\end{tabular}
\label{tab:acis}
\end{center}
Notes:\hspace{0.3cm} $^a $: Dead-time corrected exposure time.\\

\end{table}

We classified X-ray counterparts of CNe as SSSs according to their \xmm EPIC PN spectra. Spectral analysis was performed using XSPEC v12.3.1. In all our spectral fitting we used the T\"ubingen-Boulder ISM absorption (\texttt{TBabs} in XSPEC) model together with the photoelectric absorption cross-sections from \citet{1992ApJ...400..699B} and ISM abundances from \citet{2000ApJ...542..914W}. In all cases, the statistical confidence ranges of spectral parameters are 90\% unless stated otherwise. We used the $\chi^2$ fit statistic for all sources with a sufficiently high number of photon counts. For low-count spectra, the C-statistic \citep{1979ApJ...228..939C} was used and is mentioned in the text. The novae M31N~1997-11a and M31N~2006-06a were only detected or resolved in \chandra HRC-I observations. For these objects we computed HRC-I hardness ratios, as described in ``The Chandra Proposers Observatory Guide"\footnote{http://cxc.harvard.edu/proposer/POG/html/index.html;\hspace*{0.1cm} chapter 7.6}, from count rates in the bands S, M, and H (channels 1:100, 100:140, and 140:255). Additionally, for these novae we followed the method of \pz and made use of some of the \chandra ACIS-I observations mentioned above, if they were taken close to our detections. Assuming that the novae did not change X-ray brightness and spectrum between these observations, we can investigate if the corresponding X-ray source had a soft spectrum by comparing ACIS-I count rates or upper limits with HRC-I count rates. The justification for this assumption is discussed for the individual novae in Sect.\,\ref{sec:results}. While hard or moderately hard spectra lead to conversion factors (ACIS-I/HRC-I) above one, typical supersoft spectra, as found in novae, show conversion factors below 0.5.

We also conducted a general search for SSS in our \xmm data, following the approach adopted by \citet{2005A&A...434..483P} and using hardness ratios computed from count rates in energy bands 1 to 3 (0.2--0.5 keV, 0.5--1.0 keV, 1.0--2.0 keV) to classify a source. These authors defined hardness ratios and errors as

\begin{equation}
HR_i = \frac{B_{i+1} - B_i}{B_{i+1} + B_i} \; \mbox{and} \; EHR_{i} = 2 \frac{\sqrt{(B_{i+1}EB_i)^2 + (B_{i}EB_{i+1})^2}}{(B_{i+1} + B_i)^2} \; ,
\label{eqn:hardness}
\end{equation}

\noindent
for $i =$ 1,2, where $B_i$ and $EB_i$ denote count rates and corresponding errors in band $i$ as derived by \texttt{emldetect}. \pe also used these hardness ratios and classified sources as SSSs if they fulfil the conditions $HR1 < 0.0$ and $HR2 - EHR2 < -0.4$. In this work we use the same criteria.

X-ray light curves for all nova counterparts were extracted using \texttt{evselect} (XMMSAS) or \texttt{dmextract} (CIAO). A barycentre correction was always applied using \texttt{barycen} (XMMSAS) or \texttt{axbary} (CIAO). Further correction of \xmm EPIC PN source light curves was done with the XMMSAS task \texttt{epiclccorr}. For timing analysis we used the XRONOS tasks of HEASARCs software package FTOOLS \citep{1995ASPC...77..367B}\footnote{http://heasarc.gsfc.nasa.gov/ftools/}.

\section{Results}
\label{sec:results}
%
We detected eight X-ray counterparts of CNe in \m31. The positions of all objects are shown in merged images of \xmm (Fig.\,\ref{fig:xmm}) or \chandra (Fig.\,\ref{fig:chandra}). Four of these sources (see Table\,\ref{tab:novae_old_lum}) were already present in \pzk, the other four sources were detected in this monitoring for the first time (see Table\,\ref{tab:novae_new_lum}). The upper limit estimates for undetected novae from \pz are given in Table\,\ref{tab:novae_old_non}, those for undetected novae with optical outburst from May 2005 until March 2007 in Table\,\ref{tab:novae_ulim}. 

Tables\,\ref{tab:novae_old_lum} -- \ref{tab:novae_ulim} contain the following information: the name, coordinates, and outburst date of the optical nova, the distance between optical and X-ray source (if detected), the observation and its time lag to the optical outburst, the unabsorbed X-ray luminosity or the upper limit, and miscellaneous comments.

Additionally, we found four SSSs without a nova counterpart in our \xmm data. These sources are summarised in Table\,\ref{tab:sss} and are also shown in Fig.\,\ref{fig:xmm}.

\subsection{X-ray counterparts known before this work}
\label{sec:res_known}

%
\begin{table*}[ht!]
\begin{center}
\caption[]{\xmmk, \chandra and \swift measurements of \m31 optical nova candidates known from \citet{2007A&A...465..375P} and still detected here.}
\begin{tabular}{lrrlrrrl}
\hline\noalign{\smallskip}
\hline\noalign{\smallskip}
\multicolumn{3}{l}{Optical nova candidate} & \multicolumn{3}{l}{X-ray measurements} \\
\noalign{\smallskip}\hline\noalign{\smallskip}
\multicolumn{1}{l}{Name} & \multicolumn{1}{c}{RA~~~(h:m:s)$^a$} & \multicolumn{1}{c}{MJD$^b$} & \multicolumn{1}{c}{$D^c$} 
& \multicolumn{1}{c}{Observation} & \multicolumn{1}{c}{$\Delta t^d$} & \multicolumn{1}{c}{$L_{\rm X}^e$}
& \multicolumn{1}{l}{Comment$^f$} \\
M31N & \multicolumn{1}{c}{Dec~(d:m:s)$^a$} & \multicolumn{1}{c}{(d)} & (\arcsec)  & \multicolumn{1}{c}{ID} &\multicolumn{1}{c}{(d)} &\multicolumn{1}{c}{(10$^{36}$ erg s$^{-1}$)} & \\ 
\noalign{\smallskip}\hline\noalign{\smallskip}
 1996-08b & 0:42:55.2 & 50307.0 &           & 7283 (HRC-I) & 3584.3 & $< 6.9$ & \\
        & 41:20:46.0 &      & 0.9 & 0405320501 (EPIC) & 3611.6 & $3.0\pm0.6$ & SSS\\
        &      &      & 1.1 & 0405320601 (EPIC) & 3649.5 & $2.7\pm0.7$ & \\
        &      &      &           & 7284 (HRC-I) & 3701.9 & $< 13.0$ & \\
        &      &      &           & 7285 (HRC-I) & 3745.3 & $< 8.6$ & \\
        &      &      &           & 0405320701 (EPIC) & 3793.6 & $< 2.8$ & \\
        &      &      & 0.1 & 0405320801 (EPIC) & 3809.5 & $2.1\pm0.8$ & \\
        &      &      & 1.0 & 0405320901 (EPIC) & 3829.2 & $2.2\pm0.5$ & \\
        &      &      &           & 7286 (HRC-I) & 3863.6 & $< 3.7$ & \\
\noalign{\medskip}
 1997-11a & 0:42:42.13 & 50753.0 & 0.4 & 7283 (HRC-I) & 3138.3 & $9.5\pm1.4$ & SSS-HR\\
        & 41:15:10.5 &      & 0.4 & 7284 (HRC-I) & 3255.9 & $5.0\pm1.0$ & \\
        &      &      & 0.5 & 7285 (HRC-I) & 3299.3 & $7.8\pm1.6$ & \\
        &      &      & 0.3 & 7286 (HRC-I) & 3417.6 & $8.1\pm1.3$ & \\
\noalign{\medskip}
 2001-10a & 0:43:03.31 & 52186.0 &           & 7283 (HRC-I) & 1705.3 & $< 3.5$ & \\
        & 41:12:11.5 &      & 3.2 & 0405320501 (EPIC) & 1732.6 & $2.5\pm0.6$ & SSS\\
        &      &      &           & 0405320601 (EPIC) & 1770.5 & $< 5.1$ & \\
        &      &      & 0.5 & 7284 (HRC-I) & 1822.9 & $7.2\pm1.5$ & \\
        &      &      & 0.6 & 7285 (HRC-I) & 1866.3 & $11.8\pm2.0$ & \\
        &      &      & 2.0 & 0405320701 (EPIC) & 1914.6 & $3.9\pm0.8$ & \\
        &      &      & 0.9 & 0405320801 (EPIC) & 1930.5 & $3.5\pm0.7$ & \\
        &      &      &           & 0405320901 (EPIC) & 1950.2 & $< 5.2$ & \\
        &      &      & 0.9 & 7286 (HRC-I) & 1984.6 & $7.4\pm1.6$ & \\
\noalign{\medskip}
 2004-05b & 0:42:37.04 & 53143.0 &           & 7283 (HRC-I) & 748.3 & $< 2.5$ & \\
        & 41:14:28.5 &      & 1.0 & 0405320501 (EPIC) & 775.6 & $15.0\pm1.8$ & SSS\\
        &      &      & 1.3 & 0405320601 (EPIC) & 813.5 & $6.0\pm0.9$ & \\
        &      &      & & 00030802001 (XRT) & 836.7 & $3.9\pm1.2$ & \\
        &      &      & & 00030804001 (XRT) & 846.7 & $6.0\pm1.7$ & \\
        &      &      & 0.4 & 7284 (HRC-I) & 865.9 & $25.6\pm2.3$ & \\
        &      &      & 0.3 & 7285 (HRC-I) & 909.3 & $20.8\pm2.4$ & \\
        &      &      & 0.9 & 0405320701 (EPIC) & 957.6 & $23.9\pm1.3$ & \\
        &      &      & 1.6 & 0405320801 (EPIC) & 973.5 & $19.2\pm1.3$ & \\
        &      &      & 1.9 & 0405320901 (EPIC) & 993.2 & $29.8\pm1.4$ & \\
        &      &      & 0.7 & 7286 (HRC-I) & 1027.6 & $45.8\pm5.4$ & \\
\noalign{\smallskip}
\hline
\noalign{\smallskip}
\end{tabular}
\label{tab:novae_old_lum}
\end{center}
Notes: \hspace{0.2cm} $^a$: RA, Dec are given in J2000.0; $^b $: Modified Julian Date of optical outburst; MJD = JD - 2\,400\,000.5; $^c$: Distance in arcsec between optical and X-ray source; $^d $: Time after observed start of optical outburst; $^e $: unabsorbed luminosity in 0.2--10.0 keV band assuming a 50 eV blackbody spectrum with Galactic foreground absorption, luminosity errors are 1$\sigma$, upper limits are 3$\sigma$; $^f $: SSS or SSS-HR indicate X-ray sources classified as supersoft based on \xmm spectra or \chandra hardness ratios, respectively. \\
\end{table*}

From the 14 SSS nova counterparts found by \pzk, ten are no longer detected in our monitoring. Upper limits for the X-ray luminosity of these objects are given in Table\,\ref{tab:novae_old_non}. The four remaining sources are X-ray counterparts of the novae M31N~1996-08b, M31N~1997-11a, M31N~2001-10a, and M31N~2004-05b. All of them were still visible at the end of the monitoring (see Table\,\ref{tab:novae_old_lum}).

\begin{figure}
	\resizebox{\hsize}{!}{\includegraphics[angle=90]{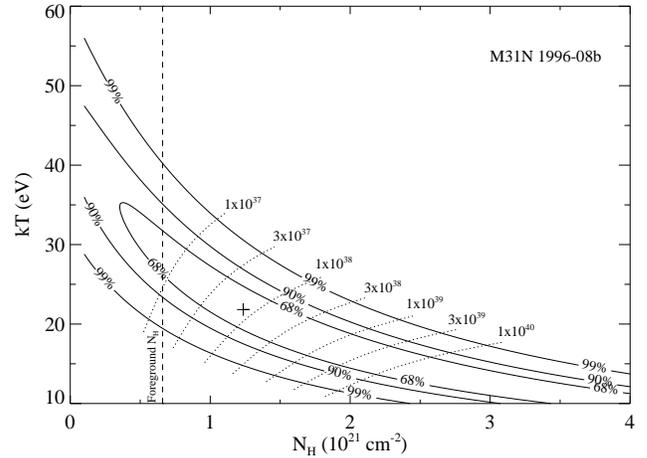}}
	\caption{Column density (\nh) - temperature (kT) contours inferred from the fit to the \xmm EPIC PN spectra of M31N~1996-08b. Indicated are the formal best-fit parameters (\textbf{plus sign}), the lines of constant X-ray luminosity (0.2-10.0 keV, \textbf{dotted lines}), and the Galactic foreground absorption (\textbf{dashed line}).}
	\label{fig:spec_n9608b}
\end{figure}

Nova M31N~1996-08b is still visible 10.6 years after the optical outburst with an X-ray luminosity that agrees with the luminosity of the last detection in February 2005 (see \pzk). Merging of all five \xmm EPIC PN observations allowed us to extract a spectrum of the X-ray source, which was previously only classified by \pz based on hardness-ratios. This spectrum does not show emission above 0.7 keV. Therefore, we classify the object as a SSS. The nominal best-fit parameters (using the C statistic) for an absorbed blackbody model are an effective temperature $kT =$ $22^{+31}_{-15}$ eV and an absorption \nh = ($1.2^{+1.8}_{-1.2}$) \hcm{21}. This results in an unabsorbed 0.2--10.0 keV band luminosity of $L_x = 7.7$\ergs{37} and a bolometric luminosity $L_{bol} = 4.4$\ergs{39}. Confidence contours for absorption column density and blackbody temperature are shown in Fig.\,\ref{fig:spec_n9608b}.

Nova M31N~1997-11a is only detected in our \chandra HRC-I observations. This is because the object is close to the centre of \m31, which causes source confusion in the \xmm images. This source is still active 9.4 years after the optical outburst. The X-ray luminosity did increase by about a factor of two with respect to the last detection in February 2005 (see \pzk). HRC-I hardness ratios $\log(S/M) = -0.10\pm0.29$ and $log(M/H) = 0.24\pm0.33$ indicate a SSS spectrum. The source is not detected in a 23 ks \chandra ACIS-I observation on 2006-12-04.88 UT (ObsID 7064, PI: S. Murray) with an upper limit of 3.2\cts{-4}. Compared with an average \chandra HRC-I count rate of about 1.4\cts{-3}, inferred from the merged HRC-I data of all four observations, this leads to an ACIS-I/HRC-I count rate factor of about 0.23. Even if we assume a decrease in count rate by a factor of two from average during the ACIS-I observation, which is a stronger variability than the source light curve in Table\,\ref{tab:novae_old_lum} indicates, the ACIS-I/HRC-I factor would still be below 0.5. This is another strong indication of a supersoft spectrum. Therefore, \chandra data provide evidence to classify nova M31N~1997-11a as a SSS.

\begin{figure}
	\resizebox{\hsize}{!}{\includegraphics[angle=90]{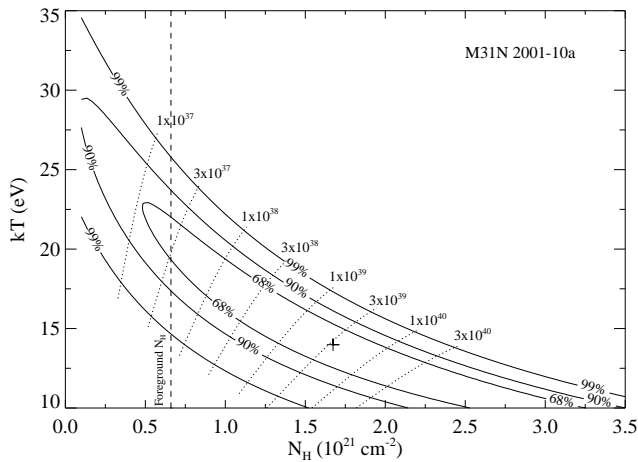}}
	\caption{Same as Fig.\,\ref{fig:spec_n9608b} for M31N~2001-10a.}
	\label{fig:spec_n0110a}
\end{figure}

Nova M31N~2001-10a is still detected 5.4 years after the optical outburst. The average luminosity of the X-ray source increased by about a factor of two with respect to the average value of \pzk, but the luminosities of the individual observations (see Table\,\ref{tab:novae_old_lum}) vary by about the same factor. Whereas \pz did not have enough information to constrain the X-ray spectrum of this source, we were now able to classify it as a SSS based on merged EPIC PN spectra of all five \xmm observations. A blackbody fit gives a best fit (C statistic) with a $kT = 14^{+15}_{-8}$ eV and a \nh = ($1.7^{+1.9}_{-1.7}$) \hcm{21}, resulting in a nominal unabsorbed $L_x = 3.3$\ergs{39} and $L_{bol} = 1.0$\ergs{43}. Figure\,\ref{fig:spec_n0110a} shows the confidence contours for absorption column density and blackbody temperature.

\begin{figure}
	\resizebox{\hsize}{!}{\includegraphics[angle=90]{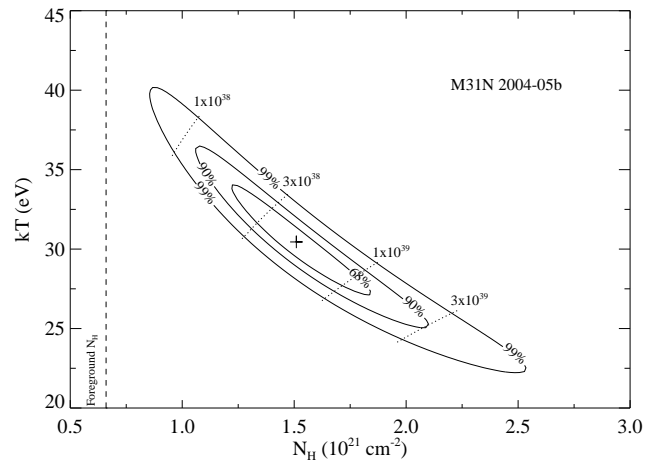}}
	\caption{Same as Fig.\,\ref{fig:spec_n9608b} for M31N~2004-05b.}
	\label{fig:spec_n0405b}
\end{figure}

Nova M31N~2004-05b is still bright in X-rays 2.8 years after the optical outburst, with an average luminosity slightly higher than the one of the last detection in February 2005 by \pzk. It is the only nova counterpart discussed here that is detected in the archival \swift XRT observations summarised in Table\,\ref{tab:acis}. The X-ray light curve during the monitoring shows strong variations by more than a factor of ten (see Table\,\ref{tab:novae_old_lum}). For the first time we are able to perform spectroscopy of this X-ray source, based on \xmm EPIC PN data. The resulting spectra can be best fitted ($\chi^2_r = 1.2$, 44 d.o.f.) with an absorbed blackbody model with $kT = 30^{+6}_{-5}$ eV and a \nh = ($1.5^{+0.5}_{-0.4}$) \hcm{21}. Therefore, we classify this source as a SSS based on its PN spectrum. Derived unabsorbed luminosities are $L_x = 4.9$\ergs{38} and $L_{bol} = 4.9$\ergs{39}. Confidence contours for absorption column density and blackbody temperature are shown in Fig.\,\ref{fig:spec_n0405b}.

\subsection{X-ray counterparts discovered in this work}
\label{sec:res_new}

\begin{table*}[ht!]
\begin{center}
\caption[]{\m31 optical novae with \xmm and \chandra counterparts discovered in this work.}
\begin{tabular}{lrrlrrrl}
\hline\noalign{\smallskip}
\hline\noalign{\smallskip}
\multicolumn{3}{l}{Optical nova candidate} & \multicolumn{3}{l}{X-ray measurements} \\
\noalign{\smallskip}\hline\noalign{\smallskip}
\multicolumn{1}{l}{Name} & \multicolumn{1}{c}{RA~~~(h:m:s)$^a$} & \multicolumn{1}{c}{MJD$^b$} & \multicolumn{1}{c}{$D^c$} 
& \multicolumn{1}{c}{Observation$^d$} & \multicolumn{1}{c}{$\Delta t^e$} & \multicolumn{1}{c}{$L_{\rm X}^f$}
& \multicolumn{1}{l}{Comment} \\
M31N & \multicolumn{1}{c}{Dec~(d:m:s)$^a$} & \multicolumn{1}{c}{(d)} & (\arcsec) & \multicolumn{1}{c}{ID} &\multicolumn{1}{c}{(d)} &\multicolumn{1}{c}{(10$^{36}$ erg s$^{-1}$)} & \\ 
\noalign{\smallskip}\hline\noalign{\smallskip}
 2001-01a & 0:42:21.49 & 51928.8 &           & mrg (HRC-I) & 1962.5 & $< 7.3$ & \\
        & 41:07:47.4 &      & 2.5 & mrg (EPIC) & 1989.9 & $2.4\pm0.5$ & SSS-HR\\
\noalign{\medskip}
 2005-02a & 0:42:52.79 & 53419.8 & 0.1 & 7283 (HRC-I) & 471.5 & $13.2\pm1.7$ & SSS\\
        & 41:14:28.9 &      & 0.1 & 0405320501 (EPIC) & 498.9 & $17.6\pm1.2$ & \\
        &      &      & 0.4 & 0405320601 (EPIC) & 536.8 & $28.4\pm1.6$ & \\
        &      &      & 0.4 & 7284 (HRC-I) & 589.1 & $30.2\pm2.8$ & \\
        &      &      & 0.4 & 7285 (HRC-I) & 632.5 & $56.4\pm4.9$ & \\
        &      &      &  & 7064 (ACIS-I) & 654.2 & $18.9\pm6.6$ & \\
        &      &      & 1.2 & 0405320701 (EPIC) & 680.9 & $20.3\pm1.3$ & \\
        &      &      & 1.1 & 0405320801 (EPIC) & 696.7 & $18.6\pm1.3$ & \\
        &      &      & 0.8 & 0405320901 (EPIC) & 716.4 & $8.9\pm0.9$ & \\
        &      &      & 0.6 & 7286 (HRC-I) & 750.9 & $23.3\pm3.9$ & \\
\noalign{\medskip}
 2006-04a & 0:43:13.42 & 53851.2 &           & 7283 (HRC-I) & 40.0 & $< 0.6$ & \\
        & 41:16:58.9 &      &           & 0405320501 (EPIC) & 67.4 & $< 0.3$ & \\
        &      &      & 1.0 & 0405320601 (EPIC) & 105.3 & $48.9\pm2.0$ & SSS, short-term variable\\
        &      &      &           & 7284 (HRC-I) & 157.6 & $< 2.0$ & \\
        &      &      &           & 7285 (HRC-I) & 201.0 & $< 7.0$ & \\
        &      &      &           & 0405320701 (EPIC) & 249.4 & $< 1.2$ & \\
        &      &      &           & 0405320801 (EPIC) & 265.2 & $< 0.5$ & \\
        &      &      &           & 0405320901 (EPIC) & 284.9 & $< 1.5$ & \\
        &      &      &           & 7286 (HRC-I) & 319.4 & $< 1.7$ & \\
\noalign{\medskip}
 2006-06a & 0:43:11.81 & 53877.0 &           & 7283 (HRC-I) & 14.3 & $< 0.7$ & \\
        & 41:13:44.7 &      &           & 0405320501 (EPIC) & 41.6 & $< 1.4$ & \\
        &      &      &           & 0405320601 (EPIC) & 79.5 & $< 1.5$ & \\
        &      &      & 0.7 & 7284 (HRC-I) & 131.9 & $7.5\pm1.8$ & SSS candidate\\
        &      &      & 0.5 & 7285 (HRC-I) & 175.3 & $10.3\pm2.1$ & \\
        &      &      &           & 0405320701 (EPIC) & 223.6 & $< 1.8$ & \\
        &      &      &           & 0405320801 (EPIC) & 239.5 & $< 1.6$ & \\
        &      &      &           & 0405320901 (EPIC) & 259.2 & $< 1.7$ & \\
        &      &      &           & 7286 (HRC-I) & 293.6 & $< 6.2$ & \\
\noalign{\smallskip}
\hline
\noalign{\smallskip}
\end{tabular}
\label{tab:novae_new_lum}
\end{center}
Notes: \hspace{0.2cm} $^a$: RA, Dec are given in J2000.0; $^b$: Modified Julian Date; MJD = JD - 2\,400\,000.5; $^c$: Distance in arcsec between optical and X-ray source; $^d$: mrg (HRC-I/EPIC) indicates the merged data of all HRC-I/EPIC observations of the monitoring; $^e$: Time after observed start of optical outburst; $^f$: unabsorbed 3$\sigma$ upper limits in 0.2--10.0 keV band assuming a 50 eV blackbody spectrum with Galactic foreground absorption.\\
\end{table*}

\subsubsection{M31N~2001-01a}
This nova candidate was discovered in the context of the Wendelstein Calar Alto Pixellensing Project (WeCAPP, \citet{2001A&A...379..362R}, see also \pzk) on 2001-01-19. In X-rays, there is a faint counterpart (see Table\,\ref{tab:novae_new_lum}), which was only found in the merged \xmm data. Because the source was not detected in any of the single \xmm observations, we assume that it was active during the whole monitoring and only the merged data provided a signal-to-noise ratio sufficient for detection. We could not construct a source spectrum because of the low count rate and the existence of a neighbouring source \citep[source 258 from ][]{2005A&A...434..483P}. However, the hardness-ratios of the source allow us to classify it as a SSS based on the criterion given in Eq.\,\ref{eqn:hardness} (Sect.\,\ref{sec:obs}).

We checked the merged \chandra HRC-I data used in \pz, which combines four archival 50 ks observations, but no counterpart of M31N~2001-01a is detected with an upper limit of $\sim$ 3\ergs{36}. This is comparable to the luminosity of our detection in this work, therefore we cannot say for sure if the source was inactive during the last observations or if it was too faint to be detected. We assume an upper limit for the turn-on time of 1990 days (first \xmm observation) and a lower limit for the turn-off time of 2317 days (last \xmm observation) for the statistics in Table\,\ref{tab:novapar}.

\subsubsection{M31N~2005-02a}
This nova candidate was discovered in the WeCAPP survey \citep{2001A&A...379..362R} on 2005-02-18. An X-ray counterpart was discovered in our first \chandra observation, 472 days after the optical outburst. The source was visible for the entire monitoring, which ended 751 days after the optical outburst. It is the only nova counterpart discussed in this work that is visible in one of the archival \chandra ACIS-I observations listed in Table\,\ref{tab:acis}. The 23 ks observation 7064 contains about seven (supersoft) source counts, implying an unabsorbed luminosity compatible with the following \xmm observation (see Table\,\ref{tab:novae_new_lum}).

\begin{figure}
	\resizebox{\hsize}{!}{\includegraphics[angle=90]{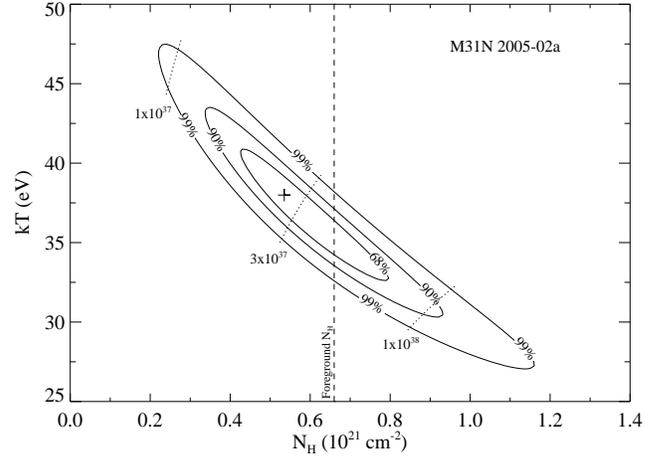}}
	\caption{Same as Fig.\,\ref{fig:spec_n9608b} for M31N~2005-02a.}
	\label{fig:spec_n0502a}
\end{figure}

The combined \xmm EPIC PN spectra can be best fitted ($\chi^2_r = 1.2$, 26 d.o.f.) with an absorbed blackbody model with $kT = 38^{+6}_{-8}$ eV and \nh = ($0.5^{+0.4}_{-0.2}$) \hcm{21}. Therefore, we classify this source as a SSS based on its PN spectrum. Note that the formal best-fit \nh is smaller than the Galactic foreground extinction of $\sim 6.7$ \hcm{20} (but still compatible with it within the errors). From the best-fit parameters we derive an unabsorbed luminosity for the 0.2--10.0 keV band $L_x = 2.6$\ergs{37} and a bolometric luminosity $L_{bol} = 1.2$\ergs{38}. Confidence contours for absorption column density and blackbody temperature are shown in Fig.\,\ref{fig:spec_n0502a}.

%
%

\begin{table*}[ht!]
\begin{center}
\caption[]{Upper limits for non-detected \m31 CN from \pzk.}
\begin{tabular}{lrrlrrrl}
\hline\noalign{\smallskip}
\hline\noalign{\smallskip}
\multicolumn{3}{l}{Optical nova candidate} & \multicolumn{3}{l}{X-ray measurements} \\
\noalign{\smallskip}\hline\noalign{\smallskip}
\multicolumn{1}{l}{Name} & \multicolumn{1}{c}{RA~~~(h:m:s)$^a$} & \multicolumn{1}{c}{MJD$^b$} & \multicolumn{1}{c}{Observation$^c$} & \multicolumn{1}{c}{$\Delta t^d$} & \multicolumn{1}{c}{$L_{\rm X}^e$}
& \multicolumn{1}{l}{Comment} \\
M31N & \multicolumn{1}{c}{Dec~(d:m:s)$^a$} & \multicolumn{1}{c}{(d)} & \multicolumn{1}{c}{ID} &\multicolumn{1}{c}{(d)} &\multicolumn{1}{c}{(10$^{36}$ erg s$^{-1}$)} & \\ 
\noalign{\smallskip}\hline\noalign{\smallskip}
 1995-09b & 0:42:43.1 & 49963.0 & mrg (HRC-I) & 3928.3 & $< 1.1$ & \\
        & 41:16:04.1 &      & mrg (EPIC) & 3955.6 & $< 37.1$ & close to the \m31 centre\\
\noalign{\medskip}
 1995-11c & 0:42:59.35 & 50047.8 & mrg (HRC-I) & 3843.5 & $< 0.2$ & \\
        & 41:16:43.1 &      & mrg (EPIC) & 3870.9 & $< 0.1$ & \\
\noalign{\medskip}
 1999-10a & 0:42:49.7 & 51454.2 & mrg (HRC-I) & 2437.0 & $< 1.0$ & \\
        & 41:16:32.0 &      & mrg (EPIC) & 2464.4 & $< 5.2$ & \\
\noalign{\medskip}
 2000-07a & 0:42:43.97 & 51752.5 & mrg (HRC-I) & 2138.8 & $< 0.2$ & \\
        & 41:17:55.5 &      & mrg (EPIC) & 2166.1 & $< 0.1$ & \\
\noalign{\medskip}
 2003-11a & 0:42:53.78 & 52948.0 & mrg (HRC-I) & 943.3 & $< 0.3$ & \\
        & 41:18:46.2 &      & mrg (EPIC) & 970.6 & $< 8.6$ & \\
\noalign{\medskip}
 2003-11b & 0:43:00.76 & 52972.8 & mrg (HRC-I) & 918.5 & $< 0.2$ & \\
        & 41:11:26.9 &      & mrg (EPIC) & 945.9 & $< 0.1$ & \\
\noalign{\medskip}
 2004-06c & 0:42:49.02 & 53181.0 & mrg (HRC-I) & 710.3 & $< 0.4$ & \\
        & 41:19:17.8 &      & mrg (EPIC) & 737.6 & $< 0.4$ & \\
\noalign{\medskip}
 2004-11b & 0:43:07.45 & 53314.8 & mrg (HRC-I) & 576.5 & $< 0.2$ & \\
        & 41:18:04.6 &      & mrg (EPIC) & 603.9 & $< 0.2$ & \\
\noalign{\medskip}
 2004-11g & 0:42:52.48 & 53314.8 & mrg (HRC-I) & 576.5 & $< 0.3$ & \\
        & 41:18:00.2 &      & mrg (EPIC) & 603.9 & $< 0.3$ & \\
\noalign{\medskip}
 2004-11e & 0:43:31.85 & 53338.8 & mrg (HRC-I) & 552.5 & $< 1.5$ & \\
        & 41:09:42.6 &      & mrg (EPIC) & 579.9 & $< 0.2$ & \\
\noalign{\smallskip}
\hline
\noalign{\smallskip}
\end{tabular}
\label{tab:novae_old_non}
\end{center}
Notes: \hspace{0.2cm} $^a$: RA, Dec are given in J2000.0; $^b$: Modified Julian Date; MJD = JD - 2\,400\,000.5; $^c$: mrg (HRC-I/EPIC) indicates the merged data of all HRC-I/EPIC observations of the monitoring; $^d$: Time after observed start of optical outburst; $^e$: unabsorbed 3$\sigma$ upper limits in 0.2--10.0 keV band assuming a 50 eV blackbody spectrum with Galactic foreground absorption.\\
\end{table*}
%

\begin{table*}[ht!]
\begin{center}
\caption[]{Upper limits for \m31 CN with outburst from about one year before the start of our monitoring till the end of our monitoring.}
\begin{tabular}{lrrrrrl}
\hline\noalign{\smallskip}
\hline\noalign{\smallskip}
\multicolumn{3}{l}{Optical nova candidate} & \multicolumn{4}{l}{X-ray measurements} \\
\noalign{\smallskip}\hline\noalign{\smallskip}
\multicolumn{1}{l}{Name} & \multicolumn{1}{c}{RA~~~(h:m:s)$^a$} & \multicolumn{1}{c}{MJD$^b$}
& \multicolumn{1}{c}{Observation$^c$} & \multicolumn{1}{c}{$\Delta t^d$} & \multicolumn{1}{c}{$L_{\rm X}^e$}
& \multicolumn{1}{l}{Comment$^f$} \\
M31N & \multicolumn{1}{c}{Dec~(d:m:s)$^a$} & \multicolumn{1}{c}{(d)} & \multicolumn{1}{c}{ID} &\multicolumn{1}{c}{(d)} &\multicolumn{1}{c}{(10$^{36}$ erg s$^{-1}$)} & \\ 
\noalign{\smallskip}\hline\noalign{\smallskip}
 2005-05a & 0:42:54.84 & 53506.0 & mrg (HRC-I) & 385.3 & $< 0.5$ & \\
        & 41:16:51.5 &      & mrg (EPIC) & 412.6 & $< 0.2$ & \\
\noalign{\medskip}
 2005-05b & 0:42:47.15 & 53532.2 & mrg (HRC-I) & 359.0 & $< 0.2$ & \\
        & 41:15:35.7 &      & mrg (EPIC) & 386.4 & $< 37.6$ & near source [PFH2005]~338\\
\noalign{\medskip}
 2005-06a & 0:42:28.42 & 53539.2 & mrg (HRC-I) & 352.0 & $< 1.9$ & \\
        & 41:16:50.9 &      & mrg (EPIC) & 379.4 & $< 0.3$ & \\
\noalign{\medskip}
 2005-06c & 0:42:31.39 & 53544.2 & mrg (HRC-I) & 347.0 & $< 3.4$ & \\
        & 41:16:20.7 &      & mrg (EPIC) & 374.4 & $< 28.5$ & near source [PFH2005]~275\\
\noalign{\medskip}
 2005-07a & 0:42:50.79 & 53581.0 & mrg (HRC-I) & 310.3 & $< 0.2$ & \\
        & 41:20:39.8 &      & mrg (EPIC) & 337.6 & $< 0.3$ & \\
\noalign{\medskip}
 2005-09a & 0:42:52.23 & 53626.0 & mrg (HRC-I) & 265.3 & $< 0.8$ & \\
        & 41:19:59.4 &      & mrg (EPIC) & 292.6 & $< 0.1$ & \\
\noalign{\medskip}
 2005-09d & 0:42:42.11 & 53635.0 & mrg (HRC-I) & 256.3 & $< 0.1$ & \\
        & 41:14:01.2 &      & mrg (EPIC) & 283.6 & $< 0.1$ & \\
\noalign{\medskip}
 2005-10b & 0:42:42.11 & 53659.8 & mrg (HRC-I) & 231.5 & $< 0.9$ & \\
        & 41:18:00.3 &      & mrg (EPIC) & 258.9 & $< 0.1$ & \\
\noalign{\medskip}
 2006-02a & 0:42:50.68 & 53768.8 & mrg (HRC-I) & 122.5 & $< 0.3$ & \\
        & 41:15:49.9 &      & mrg (EPIC) & 149.9 & $< 41.7$ & near source [PFH2005]~352\\
\noalign{\medskip}
 2006-05a & 0:43:13.94 & 53863.0 & mrg (HRC-I) & 28.3 & $< 1.8$ & \\
        & 41:20:05.6 &      & mrg (EPIC) & 55.6 & $< 1.0$ & \\
\noalign{\medskip}
 2006-06b & 0:42:32.77 & 53869.0 & mrg (HRC-I) & 22.3 & $< 1.6$ & \\
        & 41:16:49.2 &      & mrg (EPIC) & 49.6 & $< 0.1$ & \\
\noalign{\medskip}
 2006-09a & 0:42:33.16 & 53981.5 & 7284 (HRC-I) & 27.4 & $< 2.3$ & \\
        & 41:10:06.8 &      & 7285 (HRC-I) & 70.8 & $< 4.0$ & \\
        &      &      & 0405320701 (EPIC) & 119.1 & $< 0.2$ & \\
        &      &      & 0405320801 (EPIC) & 135.0 & $< 0.7$ & \\
        &      &      & 0405320901 (EPIC) & 154.7 & $< 0.3$ & \\
        &      &      & 7286 (HRC-I) & 189.1 & $< 2.4$ & \\
\noalign{\medskip}
 2006-09b & 0:42:41.45 & 53993.0 & 7284 (HRC-I) & 15.9 & $< 1.0$ & \\
        & 41:14:44.6 &      & 7285 (HRC-I) & 59.3 & $< 0.5$ & \\
        &      &      & 0405320701 (EPIC) & 107.6 & $< 9.7$ & \\
        &      &      & 0405320801 (EPIC) & 123.5 & $< 8.1$ & \\
        &      &      & 0405320901 (EPIC) & 143.2 & $< 7.4$ & \\
        &      &      & 7286 (HRC-I) & 177.6 & $< 2.3$ & \\
\noalign{\medskip}
 2006-09c & 0:42:42.38 & 53996.2 & 7284 (HRC-I) & 12.6 & $< 2.9$ & \\
        & 41:08:45.5 &      & 7285 (HRC-I) & 56.0 & $< 1.5$ & \\
        &      &      & 0405320701 (EPIC) & 104.4 & $< 2.4$ & \\
        &      &      & 0405320801 (EPIC) & 120.2 & $< 0.6$ & \\
        &      &      & 0405320901 (EPIC) & 139.9 & $< 0.9$ & \\
        &      &      & 7286 (HRC-I) & 174.4 & $< 3.1$ & \\
\noalign{\medskip}
 2006-11b & 0:42:44.05 & 54058.0 & 0405320701 (EPIC) & 42.6 & $< 8.8$ & \\
        & 41:15:02.2 &      & 0405320801 (EPIC) & 58.5 & $< 5.3$ & \\
        &      &      & 0405320901 (EPIC) & 78.2 & $< 4.9$ & \\
        &      &      & 7286 (HRC-I) & 112.6 & $< 0.7$ & \\
\noalign{\medskip}
 2006-11a$^*$ & 0:42:56.81 & 54063.8 & 405320701 (EPIC) & 36.9 & $< 0.6$ & found as SSS later\\
        & 41:06:18.4 &      & 0405320801 (EPIC) & 52.7 & $< 0.4$ & \\
        &      &      & 0405320901 (EPIC) & 72.4 & $< 0.5$ & \\
        &      &      & 7286 (HRC-I) & 106.9 & $< 12.3$ & here strongly off-axis\\
\noalign{\smallskip}
\hline
\noalign{\smallskip}
\label{tab:novae_ulim}
\end{tabular}
\end{center}
\normalsize
\vspace{0.2cm}
\end{table*}
%

\begin{table*}
\begin{center}
\addtocounter{table}{-1}
\caption[]{continued.}
\begin{tabular}{lrrrrrl}
\hline\noalign{\smallskip}
\hline\noalign{\smallskip}
\multicolumn{3}{l}{Optical nova candidate}
& \multicolumn{4}{l}{X-ray measurements}\\
\noalign{\smallskip}\hline\noalign{\smallskip}
\multicolumn{1}{l}{Name} & \multicolumn{1}{c}{RA~~~(h:m:s)$^a$} & \multicolumn{1}{c}{MJD$^b$}
& \multicolumn{1}{c}{Observation$^c$} & \multicolumn{1}{c}{$\Delta t^d$} & \multicolumn{1}{c}{$L_{\rm X}^e$}
& \multicolumn{1}{l}{Comment} \\
M31N & \multicolumn{1}{c}{Dec~(d:m:s)$^a$} & \multicolumn{1}{c}{(d)} & 
\multicolumn{1}{c}{ID} & \multicolumn{1}{c}{(d)} &\multicolumn{1}{c}{(10$^{36}$ erg s$^{-1}$)} &\\
\noalign{\smallskip}\hline\noalign{\smallskip}
 2006-12a & 0:42:21.09 & 54085.0 & 0405320701 (EPIC) & 15.6 & $< 2.7$ & \\
        & 41:13:45.3 &      & 0405320801 (EPIC) & 31.5 & $< 1.4$ & \\
        &      &      & 0405320901 (EPIC) & 51.2 & $< 6.9$ & \\
        &      &      & 7286 (HRC-I) & 85.6 & $< 4.8$ & \\
\noalign{\medskip}
 2006-12b & 0:42:11.14 & 54092.8 & 0405320701 (EPIC) & 7.9 & $< 0.3$ & \\
        & 41:07:43.8 &      & 0405320801 (EPIC) & 23.7 & $< 1.0$ & \\
        &      &      & 0405320901 (EPIC) & 43.4 & $< 0.5$ & \\
        &      &      & 7286 (HRC-I) & 77.9 & $< 2.9$ & \\
\noalign{\medskip}
 2006-12c & 0:42:43.27 & 54093.8 & 0405320701 (EPIC) & 6.9 & $< 0.5$ & \\
        & 41:17:48.1 &      & 0405320801 (EPIC) & 22.7 & $< 0.6$ & \\
        &      &      & 0405320901 (EPIC) & 42.4 & $< 0.5$ & \\
        &      &      & 7286 (HRC-I) & 76.9 & $< 0.3$ & \\
\noalign{\medskip}
 2006-12d & 0:42:44.08 & 54095.8 & 0405320701 (EPIC) & 4.9 & $< 8.7$ & probably re-brightening\\
        & 41:15:02.1 &      & 0405320801 (EPIC) & 20.7 & $< 5.3$ & of M31N~2006-11b\\
        &      &      & 0405320901 (EPIC) & 40.4 & $< 4.9$ & \\
        &      &      & 7286 (HRC-I) & 74.9 & $< 0.7$ & \\
\noalign{\medskip}
 2007-01a & 0:42:51.13 & 54114.8 & 0405320801 (EPIC) & 1.7 & $< 5.6$ & \\
        & 41:14:33.1 &      & 0405320901 (EPIC) & 21.4 & $< 6.1$ & \\
        &      &      & 7286 (HRC-I) & 55.9 & $< 0.9$ & \\
\noalign{\medskip}
 2007-02c & 0:42:39.96 & 54140.8 & 7286 (HRC-I) & 29.9 & $< 1.7$ & \\
	& 41:17:21.9 & & & & & \\
\noalign{\medskip}
 2007-03a & 0:42:53.60 & 54163.8 & 7286 (HRC-I) & 6.9 & $< 2.5$ & \\
	& 41:12:09.8 & & & & & \\
\noalign{\smallskip}
\hline
\noalign{\smallskip}
\end{tabular}
\end{center}
Notes: \hspace{0.2cm} $^{a-e}$: As in Table\,\ref{tab:novae_new_lum}; $^f$: Names of nearby sources from the catalogue of \citet{2005A&A...434..483P} ; $^*$: M31N~2006-11a was found as a SSS in \swift observations on 2007-06-01 \citep[188 d after outburst;][]{2008A&A...489..707V}\\
\end{table*}

\subsubsection{M31N~2006-04a}

%
\begin{figure}[ht!]
	\resizebox{\hsize}{!}{\includegraphics[angle=90]{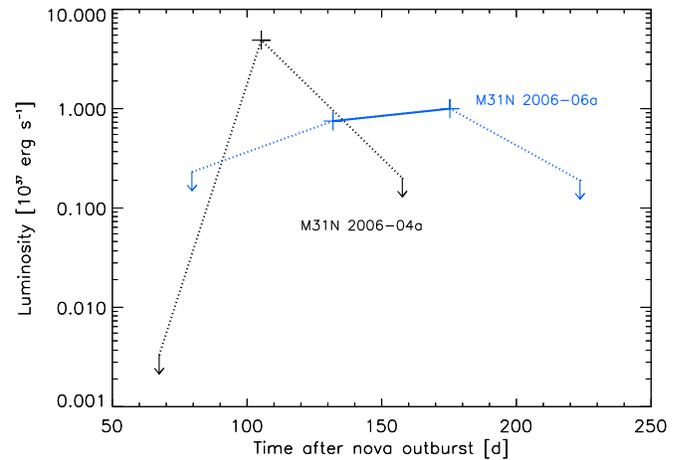}}
	\caption{Long-term X-ray light curves of the two CN counterparts with short X-ray phases presented in this work: M31N~2006-04a (\textbf{black}) and M31N~2006-06a (\textbf{blue}). Detections are marked by \textbf{crosses}. Upper limits are indicated by \textbf{arrows}. Error bars are smaller than the size of the symbols. Solid lines connect detections and dotted lines connect upper limits to detections.}
	\label{fig:xray_light}
\end{figure}
%

%
\begin{figure}[ht!]
	\resizebox{\hsize}{!}{\includegraphics[angle=90]{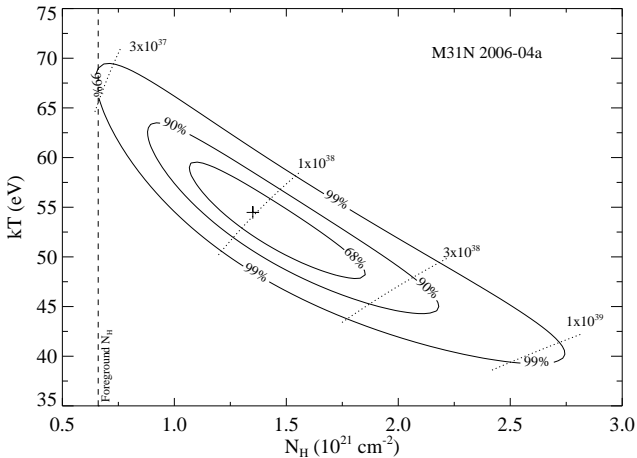}}
	\caption{Same as Fig.\,\ref{fig:spec_n9608b} for M31N~2006-04a.}
	\label{fig:spec_n0604a}
\end{figure}

This source was already discussed briefly in \citet{2010AN....331..193H} and is examined in detail here. The optical nova candidate was detected independently in the context of our optical monitoring of \m31 \citep[with the Bradford Robotic Telescope Galaxy at the Tenerife Observatory; see][]{2006ATel..805....1P} and by K. Itagaki\footnote{see http://www.cfa.harvard.edu/iau/CBAT\_M31.html\#2006-04a}. The discovery was on 2006-04-26 shortly after the \m31 visibility window opened again.

%
\begin{figure}[ht]
	\resizebox{\hsize}{!}{\includegraphics[angle=270]{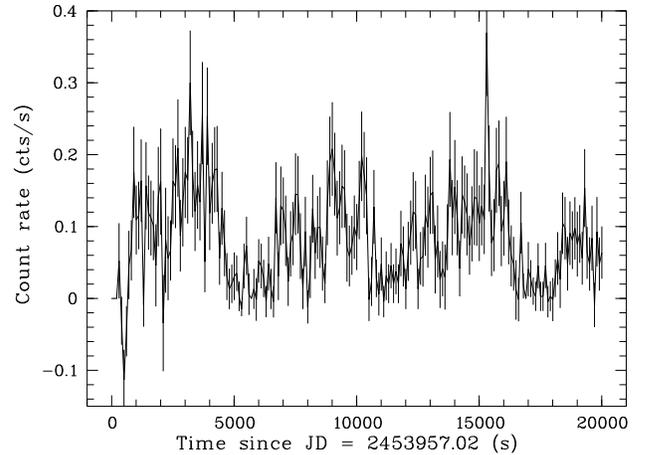}}
	\caption{Exposure and barycentre corrected short-term X-ray light curve of M31N~2006-04a (0.2 -- 1.0 keV, 100s binning) with superimposed count rate error bars.}
	\label{fig:lc_n0604a}
\end{figure}
%

%
\begin{figure}[ht!]
	\resizebox{\hsize}{!}{\includegraphics[angle=90]{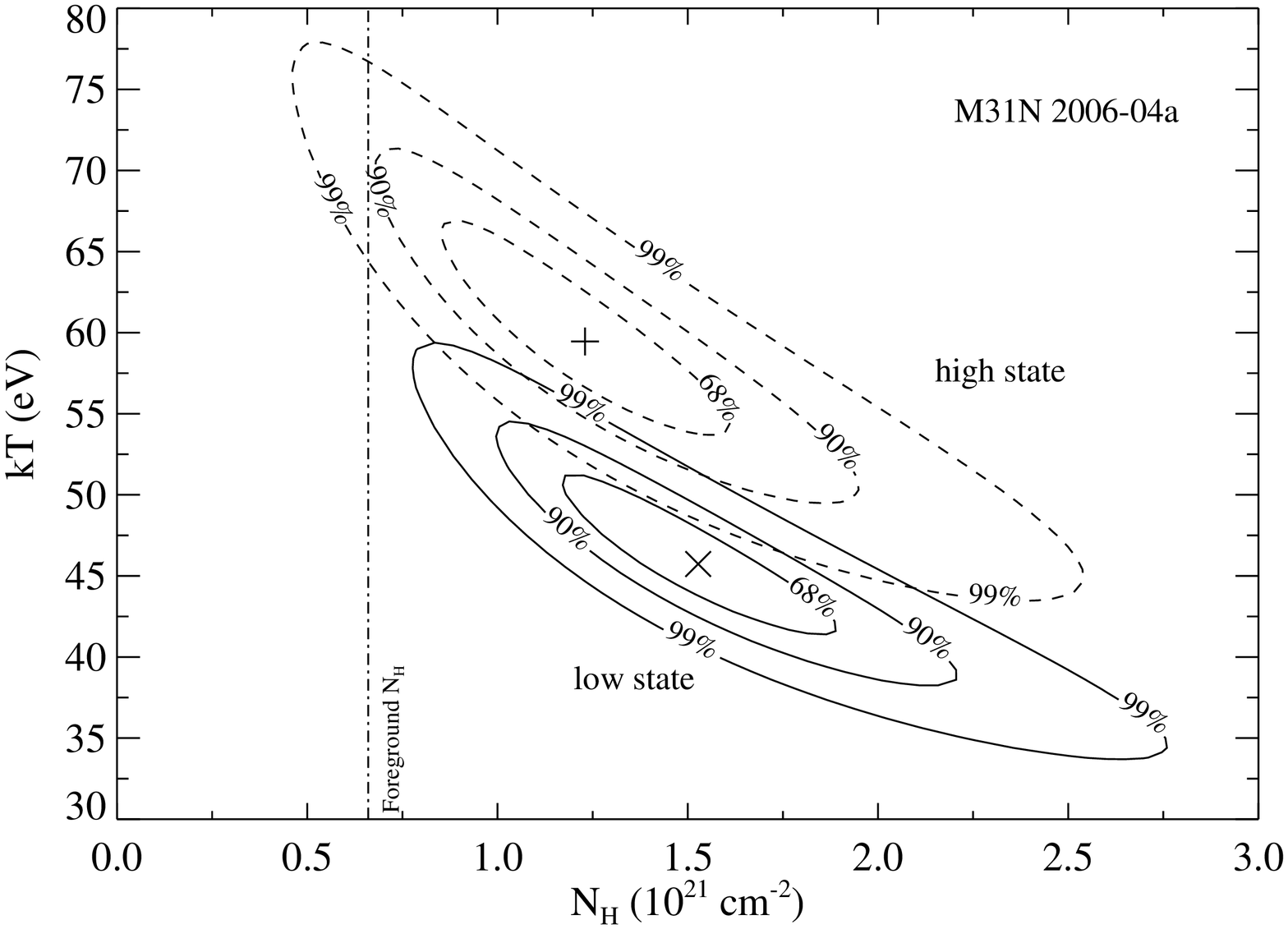}}
	\caption{Same as Fig.\,\ref{fig:spec_n9608b} for the high- and low-luminosity states of M31N~2006-04a (\textbf{dashed} and \textbf{solid} lines). Indicated are the formal best-fit parameters (\textbf{plus sign}/\textbf{cross} for high/low luminosity) and the Galactic foreground absorption (\textbf{dash-dotted line}).}
	\label{fig:spec_n0604a_hl}
\end{figure}

The X-ray counterpart was only detected in an \xmm observation 105 days after the first optical detection (see Table\,\ref{tab:novae_new_lum}). In the \swift observation 00030802001, 23 days later (see Table\,\ref{tab:acis}) on day 128, the source is not detected, but its position is close to the edge of the XRT field of view. The field of view of the second \swift observation, 10 days later, is slightly shifted and does not contain the nova position at all. From 2 or 3 photons visible in the observation on day 128 we compute a luminosity upper limit of 2.5\ergs{37}, assuming a 50 eV blackbody spectrum with Galactic foreground absorption. Because this is well above the luminosity detected from other novae in this work, the upper limit does not tell us if the SSS phase of the nova already ended at this day. The source position at the detector edge introduces further uncertainties. We therefore decided not to include this measurement in our analysis. We show the X-ray light curve in Fig.\,\ref{fig:xray_light}.

The \xmm EPIC PN spectrum of the source can be best fitted (C statistic) with an absorbed blackbody with $kT = 54^{+9}_{-10}$ eV and \nh = ($1.3^{+0.8}_{-0.5}$) \hcm{21}, allowing us to classify it as a SSS. This fit implies unabsorbed luminosities $L_x = 1.0$\ergs{38} and $L_{bol} = 2.1$\ergs{38}. Figure\,\ref{fig:spec_n0604a} shows the confidence contours for absorption column density and blackbody temperature.

The \xmm EPIC PN light curve of M31N~2006-04a during observation 0405320601, presented in Fig.\,\ref{fig:lc_n0604a}, shows variability. An FFT analysis of the light curve, performed with the FTOOLS task \texttt{powspec}, yielded a significant peak in the power spectrum towards lower frequencies. Using the FTOOLS periodicity search algorithm \texttt{efsearch}, we determined a possible period of $5900\pm900$ s ($1.6\pm0.3$ h). Note that this is about one third of the duration of our observation (20 ks). We therefore cannot place stronger constraints on the suspected periodicity. 

On the basis of the light curve we define high- and low-luminosity states for nova M31N~2006-04a, with the border between both regimes being at $\sim0.15$ cts s$^{-1}$ in Fig.\,\ref{fig:lc_n0604a}. In Fig.\,\ref{fig:spec_n0604a_hl} we show the confidence contour plot for individual absorbed blackbody fits to both states. The plot illustrates that both models are different at the 90\% confidence level, which might indicate a variation in source temperature between the two states. However, the 99\% confidence contours already overlap and the model-independent hardness ratio (0.2--0.5 vs. 0.5--0.8 keV band) supports the difference between the two states only on the 68\% level. Therefore, our analysis does not provide strong evidence for an actually variable source temperature. This caveat holds even more because blackbody models are only a first approximation to SSS spectra. The variability of M31N~2006-04a could also have been caused by variable absorption features that are not taken into consideration in our models. An example for this kind of behaviour can be found in \citet{2007ApJ...665.1334N}, who use a changing optical depth of the O\,{\sc i} absorption edge to explain variability in the SSS counterpart of the Galactic nova RS~Oph.

\subsubsection{M31N~2006-06a}
This nova candidate was discovered by \citet{2006ATel..821....1L} on 2006-05-22 and classified as a Fe II nova by \citet{2006ATel..850....1P}. An X-ray counterpart was only found in two \chandra HRC-I observations. The X-ray light curve is shown in Fig.\,\ref{fig:xray_light}. The source was not detected in a 4.1 ks \chandra ACIS-I observation on 2006-09-24.76 UT (ObsID 7140, PI: M. Garcia) with an upper limit of 1.1\cts{-3}. If we assume that the source did not change its X-ray spectrum and brightness between this observation, which is 126 days after the optical outburst, and the HRC-I observation 7284 only six days later, we derive an ACIS-I/HRC-I count rate factor upper limit of about 0.8, which is higher than 0.5 and therefore does not allow a SSS classification. However, the HRC-I hardness ratios $\log(S/M) = -0.13\pm0.33$ and $\log(M/H) = 0.37\pm0.48$ are indicating a soft spectrum. Therefore, the source can be classified as a candidate SSS based on \chandra HRC-I data.

\subsection{Non-nova supersoft sources}
\label{sec:res_sss}
We also searched our \xmm data for SSSs that do not have nova counterparts. This search was based on the hardness-ratio criterion of \pek, which is described in Eq.\,\ref{eqn:hardness} in Sect.\,\ref{sec:obs}. We found four SSSs without nova counterparts. The names of these objects are given in Table\,\ref{tab:sss} and their positions are shown in Fig.\,\ref{fig:xmm}. Three sources are already known SSSs, but the fourth SSS, hereafter CXOM31~J004228.3+411449, is detected in this monitoring for the first time. 

%
\begin{table}
\begin{center}
\caption[]{Non-nova SSSs, in the monitoring.}
\begin{tabular}{rcl}
\hline\noalign{\smallskip}
\hline\noalign{\smallskip}
\multicolumn{1}{c}{Name} & Reference$^a$ & Comment\\
\noalign{\smallskip}\hline\noalign{\smallskip}
	XMMM31~J004318.8+412017 & (1,2) & strongly variable\\
	XMMM31~J004318.7+411804 & (1) & variable\\
	XMMM31~J004252.5+411541 & (1,3) & 217.7 s period\\
	CXOM31~J004228.3+411449 & (4) & new source\\
\noalign{\smallskip} \hline
\end{tabular}
\label{tab:sss}
\end{center}
Notes:\hspace{0.1cm} $^a $: (1): \citet{2008A&A...480..599S}; (2): \citet{2006ApJ...643..844O}; (3): \citet{2008ApJ...676.1218T}; (4): this work.\\
\end{table}

The source CXOM31~J004228.3+411449 exhibits strong variability. It is only detected in the first two observations of the monitoring and its luminosity drops in the 27 days from \chandra observation 7283 to \xmm observation 0405320501 by a factor of about 40, to $1.1\pm0.1$ \ergs{37}. However, it is not detected ten days earlier in the archival \chandra ACIS-I observation 7137 (see Table\,\ref{tab:acis}) with an upper limit luminosity of $8.7$ \ergs{37}, which is well below the luminosity measured for observation 7283. The \chandra HRC-I position of the source is RA = 00:42:28.31 and Dec = +41:14:48.6 (J2000). The only nearby X-ray source we found in the literature is 2XMMi~J004227.8+411454 from the \xmm 2nd Incremental Source Catalogue (2XMMi, XMM-SSC 2008) at a distance of $7\farcs4$. It was detected as a faint and spurious object ($L_x = 2.5\pm0.9$\ergs{36}, detection likelihood 8.5) in \xmm observation 0202230401 on 2004-07-19.07 UT. However, we re-analysed the data of this observation and cannot confirm this object. This is likely because of the improved screening procedures applied by us (see Sect.\,\ref{sec:obs}) compared to the standard reduction chain used for the 2XMMi catalogue by \citet{2009A&A...493..339W}. 

We classify CXOM31~J004228.3+411449 therefore as a new source in \m31. There is no bright optical object near this position in the Local Group Galaxies Survey catalogue \citep[LGGS;][]{2006AJ....131.2478M} and there is no known nearby nova. The strong variability of CXOM31~J004228.3+411449 could indicate a very soft black hole transient, a transient SSS, or a SSS counterpart of an undetected nova. In the latter case, the fast evolution of the X-ray light curve might imply a scenario similar to M31N~2006-04a, with an optical nova outburst within one hundred days before the first detection of the SSS and a short SSS phase. We encourage optical observers to search their archives for a possible nova outburst in order to investigate this possibility.

\section{Discussion}
\label{sec:discuss}
%

\subsection{Long-term luminosity evolution of novae known from \pzk}
For all four CN counterparts described in Sect.\,\ref{sec:res_known} we find their luminosity to increase or stay constant in this work with respect to \pzk. Although the long SSS duration of all these novae seems to indicate a slow evolution, it is possible that the long-term luminosity behaviour is actually caused by a change in their X-ray spectrum. From theoretical considerations, the receding photosphere of a CN is expected to increase its effective temperature with time while keeping constant bolometric luminosity \citep[see e.g.][]{2005ASPC..330..265H}. The ECFs in Table\,\ref{tab:ecf} illustrate that a harder spectrum would result in a higher count rate for the same source flux. Since we have no detailed spectral information from earlier work it is not possible at this stage to make definite statements about the long-term luminosity evolution of these novae.

\subsection{Two recurrent novae with a very long SSS phase?}
Two novae are still observed as SSSs more than nine years after the optical outburst: M31N~1996-08b (10.6 y) and M31N~1997-11a (9.4 y). Interestingly, both novae were classified as RN candidates by \citet{2001ApJ...563..749S} and discussed as RNe in \pzk. If these two novae would turn out to be actual RNe, this would present a serious challenge to the models of X-ray emission in novae, because RNe are believed to contain a massive WD, which should lead to a very short SSS phase \citep{2005A&A...439.1061S,1998ApJ...503..381T}.

However, to search for RN candidates \citet{2001ApJ...563..749S} looked for positional coincidences between novae they discovered and novae from the literature, using a relatively large error box of $15\arcsec \times 12\arcsec$ size (with the long side orientated along the major axis of \m31). They consequently stress to have applied ''relatively loose selection criteria`` and that therefore ''these novae are recurrent nova \textit{candidates}``. We find large distances of both novae to their assumed recurrent counterparts of $9\,\farcs9$ for M31N~1996-08b to M31N~1925-01a \citep[][his nova 41]{1929ApJ....69..103H} and $7\,\farcs8$ for M31N~1997-11a to M31N~1982-09a \citep[][their nova 1]{1987ApJ...318..520C}. This leads us to the tentative conclusion that none of the two novae is a RN. In the case of M31N~1996-08b, the low effective temperature of $kT =$ ($22^{+31}_{-15}$) points towards a low mass WD \citep[see e.g.][]{2005A&A...439.1061S}. rather than a RN. Note however that for both novae none of the proposed historic nova counterparts has a published finding chart. Clarification about the nature of these novae can therefore only come from a re-examination of the original data, because historical novae can have relatively large positional errors \citep[see e.g.][]{2008A&A...477...67H}.

There is of course the possibility that both novae actually are RNe and that their last optical outbursts had been missed. But while on one hand this scenario would shorten the SSS durations considerably, on the other hand it would also result in extremely short recurrence times, implying massive WDs. There are two consequences of this scenario: (a) a massive WD should lead to an even shorter SSS phase of less than a year \citep[see e.g.][]{2006ApJS..167...59H}, and (b) earlier nova outbursts, following the same recurrence time scale, would far more likely have been detected in the past. None of these consequences agrees with the observational facts for both objects.

\subsection{X-ray variability of nova M31N~2006-04a}
The EPIC PN light curve of M31N~2006-04a (see Fig.\,\ref{fig:lc_n0604a}) points towards a ($1.6\pm0.3$) h period. Previously, there were only three other SSSs known in \m31 for which light curves indicated periodic variability. This small sample consists of the transient SSS XMMU~J004319.4+411758 \citep[865.5 s period;][]{2001A&A...378..800O}, the persistent supersoft source XMMU~J004252.5+411540 \citep[217.7 s;][]{2008ApJ...676.1218T}, and the SSS counterpart of the CN M31N~2007-12b \citep[1100 s;][]{2010AN....331..187P}. In \m31, M31N~2006-04a therefore is only the second nova to show a periodically variable X-ray flux at all and the only known SSS with a period longer than one hour.

Also in the Galaxy only six novae with periodic X-ray variability are known, namely V1494~Aql \citep[2500 s;][]{2003ApJ...584..448D}, RS~Oph \citep[35 s;][]{2006ATel..770....1O}, V4743~Sgr \citep[several periods, e.g. 6.7 h and $\sim$ 1300 s;][]{2006MNRAS.371..424L,2010MNRAS.tmp..685D}, V5116~Sgr \citep[2.97 h;][]{2008ApJ...675L..93S}, CSS 081007:030559+054715 \citep[1.77 d;][]{2009ATel.1942....1O} and KT~Eri \citep[35 s;][]{2010ATel.2423....1B}.

The typical orbital periods of CVs range between one and ten hours \citep{2003A&A...404..301R}. CNe show similar orbital periods \citep{2002AIPC..637....3W}. The pulsation periods of WDs in CNe are typically shorter than 1 hour \citep[see e.g.][2500 s pulsation period in nova V1494 Aql]{2003ApJ...584..448D}. Therefore, the X-ray variability of M31N~2006-04a might reveal the orbital period of the binary system. Alternatively, the observed variability could also be caused by an eclipsing accretion disk \citep[see e.g.][]{2010AN....331..201S} or might point towards the influence of strong magnetic fields in a similar way to what is known for thermonuclear bursts on neutron stars \citep{2006csxs.book..113S,2006MNRAS.373..769W}.

\begin{table*}[ht!]
\begin{center}
\caption[]{Observed and derived parameters of X-ray detected optical novae in
\m31.}
\begin{tabular}{llrllllrr}
\hline\noalign{\smallskip}
\hline\noalign{\smallskip}
\multicolumn{3}{l}{\normalsize{Optical measurements}} 
& \multicolumn{4}{l}{\normalsize{X-ray measurements}} 
& \multicolumn{2}{l}{\normalsize{Derived parameters}}\\
\noalign{\smallskip}\hline\noalign{\smallskip}
\multicolumn{1}{c}{Name$^a$} &
\multicolumn{1}{c}{Brightness$^b$} &
\multicolumn{1}{c}{$t_{\rm 2R}$ $^c$} &
\multicolumn{2}{c}{SSS phase} &
\multicolumn{1}{c}{$L_{\rm X}^{d}$}  &  \multicolumn{1}{c}{$kT_{\rm BB}^e$} &
\multicolumn{1}{c}{Ejected mass} & \multicolumn{1}{c}{Burned mass} \\
\noalign{\smallskip}
M31N& (mag Filter)& (d)  
& \multicolumn{1}{c}{Turn on (d)} 
& \multicolumn{1}{c}{Turn off (d)}
& \multicolumn{1}{c}{(10$^{36}$ erg s$^{-1}$)}
& \multicolumn{1}{c}{(eV)}
& \multicolumn{1}{c}{($10^{-5}$ \msun)} & \multicolumn{1}{c}{($10^{-6}$ \msun)}\\
\noalign{\smallskip}\hline\noalign{\smallskip}
 1995-09b  &15.6(H$\alpha$)&     &$1653\pm81$&$3656\pm273$&$16.1\pm3.6$& & $701^{+70}_{-67}$ & $6.1\pm0.5$ \\  
\noalign{\medskip}
 1995-11c  &16.3(H$\alpha$)&     &$762\pm725$&$3609\pm236$&$13.8\pm3.3$&  & $149^{+419}_{-149}$ & $6.0\pm0.4$ \\  
\noalign{\medskip}
 1996-08b$^*$  &16.1(H$\alpha$)&     &$1831\pm49$& $>3829$ &$5.6\pm1.0$&$22^{+31}_{-15}$ & $861^{+47}_{-45}$ & $>6.4$ \\
\noalign{\medskip}
 1997-11a$^*$ &18.0(R)&     &$2027\pm566$& $>3418$ &$9.5\pm1.4$& & $1060^{+670}_{-510}$ & $>5.7$ \\
\noalign{\medskip}
 1999-10a   &17.5(W)&     &$357\pm496$&$2203\pm234$&$21.2\pm1.6$&$34^{+4}_{-4}$ & $33^{+154}_{-28}$ & $3.7\pm0.4$ \\
\noalign{\medskip}
 2000-07a   &16.8(R)&22.4&$162\pm8$&$1904\pm236$&$13.4\pm0.7$& & $6.7\pm0.7$ & $3.2\pm0.4$ \\
\noalign{\medskip}
 2001-01a$^*$  &17.1(R)&     &$<1990$&$>2317$&$2.4\pm0.5$& & $<1016$ & $>3.9$ \\
\noalign{\medskip}
 2001-10a$^*$  &17.0(R)&39.3&$1089\pm70$& $>1985$ &$12.0\pm2.0$ & $14^{+15}_{-8}$ & $304^{+40}_{-38}$ & $>3.3$ \\
\noalign{\medskip}	  
 2003-11a   &16.9(R)&22.0&$327\pm70$&$709\pm235$&$27.6\pm1.8$& & $27^{+13}_{-10}$ & $1.2\pm0.4$ \\
\noalign{\medskip}					  
 2003-11b &17.4(R)&42.2&$302\pm70$&$684\pm235$&$10.4\pm1.3$& & $23^{+12}_{-10}$ & $1.1\pm0.4$ \\
\noalign{\medskip}
 2004-05b$^*$  &17.2(R)&49.7&$213\pm11$& $>1028$ & $45.8\pm5.4$&$30^{+6}_{-5}$ & $11.7\pm1.2$ & $>1.72$ \\
\noalign{\medskip}
 2004-06c  &17.1(R)&10.9&$94\pm70$&$476\pm234$&$33.9\pm2.0$& & $2.3^{+4.7}_{-2.1}$ & $0.8\pm0.4$ \\
\noalign{\medskip}
 2004-11b$^+$ &16.6(R)&32.0&$68\pm16$&$343\pm235$&$16.2\pm1.6$& & $1.9^{+1.0}_{-0.8}$ & $0.6\pm0.4$ \\
\noalign{\medskip}
 2004-11g   &18.0(R)&28.4&$16\pm16$&$343\pm235$&$82.1\pm3.0$& & $0.07^{+0.20}_{-0.07}$ & $0.6\pm0.4$ \\
\noalign{\medskip}
 2004-11e  &17.6(R)&34.6&$44\pm16$&$319\pm235$&$35.9\pm4.0$& & $0.5^{+0.4}_{-0.3}$ & $0.5\pm0.4$ \\
\noalign{\medskip}
 2005-02a$^*$  &17.7(W)&     &$236\pm236$& $>751$ &$56.4\pm4.9$&$38^{+6}_{-8}$ & $14.3^{+42.9}_{-14.3}$ & $>1.25$ \\
\noalign{\medskip}
 2006-04a$^*$  &15.9(W)&     &$86\pm19$&$132\pm27$&$48.9\pm2.0$&$54^{+9}_{-10}$ & $1.9^{+0.9}_{-0.8}$ & $0.22\pm0.05$ \\
\noalign{\medskip}
 2006-06a$^*+$ &17.6(R)&     &$106\pm26$&$200\pm25$&$10.3\pm2.1$&  & $2.1^{+1.2}_{-0.9}$ & $0.33\pm0.04$ \\
\noalign{\medskip}
\hline
\noalign{\smallskip}
\end{tabular}
\label{tab:novapar}
\end{center}
Notes:\hspace{0.2cm} $^a $: $^*$ indicates novae with parameters mainly from this paper while the remaining objects are from \pz (their table 8) and we give updated turn-off times, ejected masses and burned masses, $^+$  indicates novae confirmed by optical spectra; $^b $: maximum observed magnitude (not necessarily peak magnitude), ``W" indicates unfiltered magnitude; $^c $: time in days the nova R magnitude needs to drop 2 mag below peak magnitude \citep[see][]{1964gano.book.....P}; $^d $: unabsorbed luminosity in 0.2--1.0 keV band in units of 10$^{36}$ erg s$^{-1}$ during observed maximum X-ray brightness assuming a 50 eV blackbody spectrum with Galactic foreground absorption; $^e $: blackbody temperature from spectral fit.\\
\end{table*}

\subsection{Duration of SSS phases of novae}
\label{sec:discuss_dur}
Among the X-ray counterparts of novae discovered in this work there are two sources with short SSS phases (M31N~2006-04a and M31N~2006-06a; $46\pm46 d$ and $94\pm51 d$) and only one that is visible as a bright SSS throughout the monitoring (M31N~2005-02a). This confirms the finding of \pzk, who detected a significant number of nova counterparts with short SSS phases in their work.

\subsection{Nova parameters}
\label{sec:discuss_par}
In Table\,\ref{tab:novapar} we summarise the optical and X-ray parameters of all novae that were found in this work or that were still active in \pz but are not detected anymore. The optical information includes nova confirmation by optical spectra, observed maximum brightness and the decay time $t_{\rm 2R}$, which is defined as the time in days the nova luminosity needs to drop 2 magnitudes below peak luminosity \citep{1964gano.book.....P}. Based on the X-ray measurements, we give the turn-on and turn-off time of the SSS phase, the maximum X-ray luminosity and the temperature inferred from a blackbody fit. As mentioned above, blackbody fits to SSS spectra of novae tend to underestimate the effective temperature. This should be kept in mind when using the $kT$ values given in Table\,\ref{tab:novapar}. From the observed parameters we derived the mass ejected in the outburst and the mass burned in the remaining shell. 

The ejected mass in a nova outburst can be estimated from the turn-on time of the SSS and from the expansion velocity of the ejected material. The decrease in the optical thickness of the expanding ejecta is responsible for the rise in the X-ray light curve of the post-outburst novae, as shown for V1974 Cyg \citep{1996ApJ...463L..21S,1996ApJ...456..788K}. Following \citet{2002A&A...390..155D} we assume the volume of the nova shell, expanding at constant velocity v, to be $V \sim 4\pi \times \mbox{v}^3 \times t^3 \times f$, where the fill factor $f$ is a dimensionless parameter describing the thickness of the shell. We chose $f = 0.2$, which incorporates the increase of the thickness of the envelope due to thermal motions inside the gas \citep[see][and references therein]{2002A&A...390..155D}. Under this assumption, the column density of hydrogen evolves with time as $N_{H} ({\rm cm}^{-2})= M^{ej}_{H} / (\frac{4}{3}\pi m_H \mbox{v}^{2} t^{2} {f}')$, where $m_H=1.673\times10^{-24}$ g is the mass of the hydrogen atom and $f' \sim 2.4$ (for $f = 0.2$) a geometric correction factor. Note that \pz used a different approach to compute ejected masses and we therefore give updated values for all objects in Table\,\ref{tab:novapar}. For most novae, we assumed a typical value for the expansion velocity of 2000 km s$^{-1}$. The estimates for the ejected mass can be improved for novae with measured outflow velocities, but only for two objects these values are known from optical spectra: M31N~2004-11b (2500 km s$^{-1}$)\footnote{see Filippenko et al. at \\ http://cfa-www.harvard.edu/iau/CBAT\_M31.html} and M31~2006-06a \citep[1700 km s$^{-1}$;][]{2006ATel..850....1P}. We further assume that the SSS turns on when the absorbing hydrogen column density decreases to $\sim10^{21}$ cm$^{-2}$. 

The turn-off time of the SSS phase allows us to estimate the amount of hydrogen-rich material burned on the WD surface, $M^{burn}=(L\Delta t) / (X_H \epsilon)$, where $L$ is the bolometric luminosity, $\Delta t$ the turn-off time, $X_H$ the hydrogen fraction of the burned material, and $\epsilon=5.98\times10^{18}$ erg g$^{-1}$ \citep{2005A&A...439.1061S}. Following \pzk, we assumed a bolometric luminosity of $3\times10^4L_{\odot}$ and a hydrogen mass fraction $X_H=0.5$. Note that we do \textit{not} use the duration of the SSS phase as $\Delta t$, as was done in \pzk. The reason is that the steady nuclear burning phase already starts before the turn-on of the SSS, a time which is dependent on the ejected mass. But hydrogen burning is powering the nova since the outburst and later settles to be stable after the inner layers of the envelope returned to hydrostatic equilibrium \citep{2005A&A...439.1061S}. The onset of stable burning is not well defined theoretically but hydrogen clearly is already burned before the SSS becomes visible. Therefore, the turn-off time of the SSS provides a better base to estimate the hydrogen mass burned on the WD surface.

The hydrogen mass fraction of a nova is rarely determined observationally and the uncertainties associated with the above assumed values are not established. From theoretical models by \citet{2005A&A...439.1061S} one sees that e.g. for a high degree (75\%) of mixing of the accreted envelope with the degenerate core (which results in $X_H = 0.18$) the stable hydrogen-burning phase is expected to last less than two months, even for a small WD mass of $0.9 M_{\sun}$. Therefore, the chemical abundances of a nova strongly affect the intrinsic duration of the SSS phase and $X_H \sim 0.3$ would be a more appropriate choice for fast and bright novae. However, overall $X_H = 0.5$ can be considered as a reasonable assumption.

We decided to use a constant bolometric luminosity to compute the burned masses, because the luminosities obtained from spectral fits are partly unphysically high and possess huge errors that would dominate the resulting mass estimates. This approach allows us to concentrate on the more robustly determined source parameters. However, bolometric luminosities of novae are expected to be close to the Eddington limit. \citet{2005A&A...439.1061S} determine in their model a plateau luminosity in the range of 2--6 \tpower{4} $L_{\sun}$, which includes our generic assumption and can very well serve as its error range.

Despite the uncertainties, the burned masses presented in Table\,\ref{tab:novapar} are within the range expected from models of stable envelopes with steady hydrogen burning \citep{2005A&A...439.1061S,1998ApJ...503..381T}. In general, the burned masses are about one order of magnitude smaller than the ejected masses, which for most novae are within the values predicted from hydrodynamical models of nova outbursts \citep{1998ApJ...494..680J}. The exceptions are novae M31N~1995-09b, M31N~1996-08b, and M31N~1997-11a, for which we estimate very large masses. Above, we rejected the interpretation that the latter two objects might be recurrent novae. Moreover, that neither nova has yet reached the end of its SSS phase points towards a burning envelope with a high mass and/or a large hydrogen fraction. Because a high envelope mass points towards a low WD mass, the WD masses in these nova systems might be below the mass range used by \citet{2005A&A...439.1061S} ($0.9 M_{\sun}$) and by \citet{1998ApJ...494..680J} ($0.8 M_{\sun}$) in their models. However, without information on the degree of mixing of the accreted material with the degenerate core, we cannot draw conclusions on the actual long-term evolution of the WD mass.

A detailed statistical analysis of the nova parameters will be presented in the second paper (Henze et al., in prep.) and will include a larger sample of objects.

\subsection{Novae without detected SSS phase}
We find that out of 25 known CNe, which were in the field of view of our observations and had their outburst from about one year before the start of the first observation till the end of the monitoring, 23 were not detected in X-rays (see Table\,\ref{tab:novae_ulim}). Some of these novae have positions that are close to bright persistent X-ray sources in the field, but for all of them our upper limits rule out a luminous SSS counterpart ($L_x \gtrsim$ \oergs{37}) during the observations.

The percentage of novae with non-detected SSS phase is much higher in our monitoring than in the work of \pzk. These authors found that 21 out of 32 novae with optical outbursts from November 2003 to February 2005 were not detected in their X-ray data (see their Table 7), which covered the period from July 2004 to February 2005. We performed Monte Carlo simulations based on a preliminary empirical distribution of nova SSS phase turn-on and turn-off times. We found that both results agree within the statistical $2\sigma$ errors. The discrepancy between the reported numbers is likely caused by the distribution of the optical outburst dates of the \textit{discovered} novae.

In the second paper we will discuss the question of completeness of our monitoring in a quantitative way on the basis of a larger data set. However, we mention here three qualitative explanations for the large number of CNe with undetected SSS phase. First, some novae might actually turn on as SSS after our monitoring, implying a massive ejected envelope (see Sect.\,\ref{sec:discuss_par}). Second, a significant number of CNe are likely to show short SSS phases (see Sect.\,\ref{sec:discuss_dur}). Fast objects like nova M31N~2004-11f from \pz (SSS duration $\sim$ 30 d) might therefore still be missed by our monitoring because of the sampling of the observations. The third scenario, which has not been discussed in the literature so far, is that at the time when the ejected nova envelope becomes optically thin, the hydrogen burning in the WD atmosphere has already ceased. Therefore, these objects would be hidden from detection as SSSs by self-absorption. The exact parameters for this scenario are not known, mainly because in the case of non-hydrostatic equilibrium burning the assumptions we use to compute the burned mass are not valid. Qualitatively, strong mixing would be required to reduce the hydrogen content in the burning layer and increase the (CNO) metallicity of the ejected envelope, thus increasing its optical thickness.

\section{Summary}
\label{sec:summary}
%
In this paper we describe the first X-ray monitoring dedicated to find and examine counterparts of classical novae in \m31. The earlier studies of \pe and \pz made use of archival data. We detected eight X-ray counterparts of CNe, seven of which can be classified as SSSs, and another four SSSs without optical counterpart. We can therefore confirm the finding of \pe that novae are the major class of SSSs in the central region of \m31. Two of the X-ray counterparts are still visible more than nine years after the nova outburst, whereas two other nova counterparts show a short SSS phase of less than 150 days. One of the latter sources, M31N~2006-04a, shows a short-time variable X-ray light curve with an apparent periodicity of ($1.6\pm0.3$) h. For 23 out of 25 CNe, which were in the field of view of our observations and had their outburst from about one year before the start of the monitoring till its end, there is no X-ray counterpart detected. From the 14 SSS nova counterparts known from previous studies, ten are not detected anymore.

In a second paper (Henze et al., in prep.), we will present results from our \m31 monitoring campaigns in 2007/8 and 2008/9. In order to take into account CNe with short SSS phases these monitorings consisted of observations separated by ten days, instead of 40 days as in this work. Owing to the change in observational strategy, these data will be analysed in the second paper, which will also contain an extensive discussion of the complete sample of known CNe with SSS counterparts in \m31. 

\begin{acknowledgements}
We wish to thank the referee, Jan-Uwe Ness, for his constructive comments, which helped to improve the clarity of the manuscript. The X-ray work is based in part on observations with \xmmk, an ESA Science Mission with 
instruments and contributions directly funded by ESA Member States and NASA. The \xmm project 
is supported by the Bundesministerium f\"{u}r Wirtschaft und Technologie / Deutsches Zentrum 
f\"{u}r Luft- und Raumfahrt (BMWI/DLR FKZ 50 OX 0001) and the Max-Planck Society. M. Henze acknowledges support from the BMWI/DLR, FKZ 50 OR 0405. A.R. acknowledges support from SAO grant GO9-0024X. M. Hernanz acknowledges support from grants AYA2008-01839 and 2009-SGR-315. G.S. acknowledges support from grants AYA2008-04211-C02-01 and AYA2007-66256.

\end{acknowledgements}
\bibliographystyle{aa}

\end{document}